\documentclass[
	aps,
	showpacs,
	showkeys,
	twocolumn,
	altaffilletter,
	nolongbibliography,
	numerical,
	flushbottom,
	secnumarabic,
	pra,
	superscriptaddress,
	floatfix,
	10pt
]{revtex4-1}

\usepackage{lipsum}

\usepackage[dvipsnames]{xcolor}

\usepackage{bm}

\usepackage{graphicx}

\usepackage{upgreek}

\usepackage{natbib}

\usepackage{braket}

\usepackage{physics}

\usepackage{newtxtext, newtxmath}

\usepackage{siunitx}

\usepackage[caption=false]{subfig}
\usepackage[%
	breaklinks,%
	pdftex,%
	hyperfootnotes=true,%
	pdfpagelabels,%
	bookmarks,%
	pageanchor,%
]{hyperref}

\hypersetup{%
	colorlinks=true, linktocpage=true, pdfstartpage=1, pdfstartview=FitH, pdfborder={0 0 0},%
	breaklinks=true, pdfpagemode=UseNone, pageanchor=true, pdfpagemode=UseOutlines,%
	plainpages=false, bookmarksnumbered, bookmarksopen=true, bookmarksopenlevel=1,%
	hypertexnames=true, pdfhighlight=/O,
	urlcolor=Red!50!Sepia, linkcolor=NavyBlue, citecolor=RoyalBlue, 
}

\graphicspath{{Pictures/}}

\newcommand{\ie}{i.\,e.}

\newcommand{\cc}{\ensuremath{\text{c.\,c.}}}
\newcommand{\eg}{e.\,g.}
\newcommand{\ped}[1]{_\text{#1}}
\newcommand{\api}[1]{^\text{#1}}

\newcommand{\ham}{H}
\newcommand{\tf}{t\ped{f}}
\newcommand{\diss}{\mathcal{D}}
\newcommand{\partitionFunction}{Z}

\newcommand{\timeordering}{\mathcal{T}}
\newcommand{\DELTA}{\mathop{}\!\Updelta}

\newcommand{\rng}[2]{\ensuremath{[#1, #2]}}
\renewcommand{\epsilon}{\varepsilon}

\newcommand{\beq}{\begin{equation}}
\newcommand{\eneq}{\end{equation}}

\DeclareMathOperator{\eu}{e}
\DeclareMathOperator{\iu}{i}
\DeclareMathOperator{\Ci}{Ci}
\DeclareMathOperator{\Sf}{Sf}

\begin{document}

\author{L.~M.~Cangemi}
\email{lorismaria.cangemi@unina.it}
\author{G.~Passarelli}
\author{V.~Cataudella}
\author{P.~Lucignano}
\author{G.~De Filippis}
\affiliation{Dipartimento di Fisica ``E.~Pancini'', Universit\`a di Napoli ``Federico II'', Complesso di Monte S.~Angelo, via Cinthia, 80126 Napoli, Italy}
\affiliation{CNR-SPIN, c/o Complesso di Monte S. Angelo, via Cinthia - 80126 - Napoli, Italy}

\title{Beyond the Born-Markov approximation: dissipative dynamics of a single qubit}

\date{\today}


\begin{abstract}
	 We propose a numerical technique based on a combination of short-iterative Lanczos and exact diagonalization methods, suitable for simulating the time evolution of the reduced density matrix of a single qubit interacting with an environment. By choosing a mode discretization method and a flexible bath states truncation scheme, we are able to include in the physical description multiple-excitation processes, beyond weak coupling and Markov approximations. 
	 We apply our technique to the simulation of three different model Hamiltonians, which are relevant in the field of adiabatic quantum computation. We compare our results with those obtained on the basis of the widely used Lindblad master equation, as well as with well-known exact and approximated approaches. We show that our method is able to recover the thermodynamic behavior of the qubit-bath system, beyond the Born-Markov approximation. Finally, we show that even in the case of the adiabatic quantum annealing of a single qubit the bath can be beneficial in reaching the reduced system ground state.
\end{abstract}

\maketitle

\section{Introduction}\label{sec:intro}

	Quantum mechanical systems interacting with their surroundings experience dissipation and decoherence. The prototypical model aimed at describing an open quantum system is based on a quantum two-level system (TLS) interacting with a bosonic bath in thermal equilibrium at fixed temperature, the so-called spin-boson model (SBM)~\cite{caldeira:caldeira-leggett, breuer:open-quantum}.
	Several approaches have been proposed in order to attack this problem. The main idea is to include effectively the environmental noise in the evolution of the dynamical variables of the system of interest (reduced system). This led to successful tools for studying the single qubit open dynamics, such as real-time path-integral Monte Carlo (rt-PIMC)~\cite{PathInt1,PathInt2}, quasi-adiabatic propagators (QUAPI)~\cite{quapi, weenie-dattani, weenie-dattani-2}, non-interacting or weakly-interacting blip approximation (NIBA or WIBA)~\cite{Leggett,Grifoni}, numerical time-dependent renormalization group (NRG)~\cite{LeHurNRG1,LeHurImpurity} or quantum master equations (QME), such as the celebrated Gorini-Kossakowski-Sudarshan-Lindblad equation~\cite{lindblad1976,GoriniKossSud76} (Lindblad equation from now on for brevity). 
	
	The Lindblad equation is a consolidated tool for studying the dynamics of open quantum systems. More recently, it has been widely used to describe decoherence effects in adiabatic quantum computation (AQC) and quantum annealing (QA)~\cite{kadowaki:qa,farhi:quantum-computation,childs:robustness}, fields that regained momentum since the first experimental demonstration of the D-Wave machine~\cite{harris:d-wave}. The Lindblad equation relies on several assumptions on the system dynamics: in particular, on the Born approximation (disregarding qubit-bath correlations at any times during the dynamics provided that their coupling energy is weak enough) and on the Markov approximation (which ensures that the dissipation mechanism involves no memory effects). Moreover, it is strictly valid only in the purely adiabatic regime, where the rotating wave approximation (RWA) holds. 
	
	Recent works~\cite{Me,Smelyanskiy:decoherence,Smelyanskiy:decoherence2,amin:thermal-qa,dickson:thermal-qa,arceci:dissipative-lz} show that AQC of a qubit ensemble with intermediate coupling to its bath may have shorter annealing time than a closed system. This speed-up is predicted at very low temperatures and intermediate couplings to the environment: a regime where non-Markovian effects and multiple-excitation processes may be relevant~\cite{CialdiExp,NatureChina,thorwart3,goychuk:2006}. 

	In this work, we discuss an alternative technique to account for decoherence and dissipation in open quantum systems, which could in principle overcome the limitations of the Lindblad equation, allowing to disengage from the Born and Markov approximations. This approach combines: a discretization of the bath~\cite{de-vega:discretize-baths}, which is described in terms of a finite number of independent harmonic oscillators; a smart truncation scheme of the bosonic Hilbert space; short-iterative Lanczos (SIL) method~\cite{Lanczos1,Lanczos2,nature:giulio,letters:giulio1}. Our technique is not affected by the limitations of standard perturbative approaches, as it guarantees the trace preservation and positivity of the density operator.
	Due to its stability and reduced computational effort, this method allows to include multiple-excitation processes, not accounted by the Lindblad theory~\cite{prb:giulio, letters:giulio2}. Moreover, as we do not trace the bath degrees of freedom, we have access to the full wave function, and we can measure all the properties of either the reduced system and the bath.
	In order to test the reliability of this approach, we will focus on three models describing a TLS interacting with the environment. 
	
	This paper is organized as follows: in Sec.~\ref{sec:environment}, we introduce a general model Hamiltonian of the system we intend to study, outlining the characteristics of the dissipation; in Sec.~\ref{sec:numerics}, we discuss the main features of our numerical method; in Sec.~\ref{sec:qubits}, we introduce the particular TLS Hamiltonians to be studied; the different approximations schemes known in the literature and related results are discussed and compared. We present our results regarding the analyzed models, compare them with known approximations and finally discuss further possible extensions of this work in Sec.~\ref{sec:conclusions}.
	
	\section{Model Hamiltonian}\label{sec:environment}
	
	Our qubit is described by the time-dependent Hamiltonian $ \ham_S(t) $. The full system-environment Hamiltonian is usually written as
	\begin{equation}\label{eq:system-bath-hamiltonian}
		\ham(t) = \ham_S(t) + \ham_B + V,
	\end{equation}
	where $ V $ is a time-independent interaction potential between the two subsystems and $ \ham_B $ is the environment Hamiltonian. 
	The qubit can be viewed as an effective spin one-half particle. The eigenvectors of the Pauli operator $ \sigma_z $ represent the computational basis; the spin-flipping operators $ \sigma_\pm $ allow for quantum tunneling between these two states. In general, we take $ \ham_S(t) $ as a time-dependent real and symmetric operator of the form
	\begin{equation}\label{eq:tls-hamiltonian}
	\ham_S(t) = -\Gamma(t) \sigma_x - \epsilon(t) \sigma_z,
	\end{equation}
	where $ \Gamma(t) $ is a transverse field and $ \epsilon(t) $ fixes the energy bias between the TLS states.
	As customary, we model the environment by a collection of independent bosons, and the Hamiltonian $ \ham_B $ ($ \hslash = 1 $ here and in the following) reads
	\begin{equation}\label{eq:bath-hamiltonian}
	\ham_B = \sum_k \omega_k b_k^\dagger b_k, \qquad \comm{b_l}{b_m^\dagger} = \delta_{lm}.
	\end{equation}
	Here, $ \omega_k $ are mode frequencies and $ b_k $ ($ b_k^\dagger $) annihilates (creates) a boson in mode $ k $; we omit the zero-point energy $ \sum_k \omega_k/2 $.
	
	The environment acts locally on the qubit system, coupling to $ \sigma_z $. In the dipole approximation, qubit eigenstates are coupled to each bosonic displacement operator, and the interaction Hamiltonian reads
	\begin{equation}\label{eq:interaction-hamiltonian}
		V = \sigma_z \sum_k g_k \qty(b_k + b_k^\dagger) \equiv \sigma_z \otimes B;
	\end{equation}
	$ g_k $ is the coupling energy among the reduced system and the $ k $th bosonic mode. The dipole approximation is valid only if every $ g_k $ is weak when compared to the other energy scales~\cite{Leggett}.
	All the details concerning dissipation are contained in the bath density of states $ J(\omega) $, defined as
	\begin{equation}\label{eq:spectral-function}
		J(\omega) = \sum_k g_k^2 \delta\qty(\omega - \omega_k).
	\end{equation}
	When the density of modes is large enough, $ J(\omega) $ behaves as a continuous function, \ie, as a power-law of $ \omega $ for $ \omega\to0 $, up to a high-energy cut-off $ \omega\ped{c} $:  
	\begin{equation}\label{eq:spectral-function-continuous}
		J(\omega) = \sum_k g_k^2 \delta\qty(\omega - \omega_k) = \eta \frac{\omega^s}{\omega\ped{c}^{s-1}} \Theta(\omega\ped{c} - \omega),
	\end{equation}
	where $ \eta $ is an effective dimensionless coupling and $ \Theta(x) $ is the Heaviside step-function. The step-function could be replaced by an exponential decay or a Lorentzian tail, but in our case it is advisable to work with a sharp cut-off for reasons that will be clearer in the following. Needless to say, physical results must be independent both of $ \omega\ped{c} $ and the form of the cut-off.  The exponent $ s $ determines the nature of the dissipation: sub-Ohmic ($ 0 < s < 1 $), Ohmic ($ s = 1 $) or super-Ohmic ($ s > 1 $). In this work, we will discuss the three representative cases $ s = 1/2 $, $ s = 1 $ and $ s = 2 $. 
	
	The two-level Hamiltonian~\eqref{eq:tls-hamiltonian} can be diagonalized exactly for every choice of $ \Gamma(t) $ and $ \epsilon(t) $; however, the problem complicates enormously when the full Hamiltonian~\eqref{eq:system-bath-hamiltonian} is taken into account, and a closed-form analytical solution is not known in general. In what follows, we shall discuss a numerical approach suited for dealing with the full system dynamics governed by Eq.~\eqref{eq:system-bath-hamiltonian}. It will prove useful in studying the time-independent limits of Hamiltonian~\eqref{eq:system-bath-hamiltonian}, as well as the fully time-dependent case, where analytical solutions are not available.
	
	\section{Short-iterative Lanczos method}\label{sec:numerics}
	
	At the initial time $ t_{0} $, we assume the density matrix of the qubit and bath to be factorized:
	\beq
		\rho(t_{0})=\rho_{S}(t_{0}) \otimes \rho_{B},
	\eneq  
	where $ \rho_{S}(t_{0}) $ is the density operator of the reduced system at the initial time and $ \rho_{B} $ is the bath density operator at thermodynamic equilibrium at temperature $ T = 1/\beta $ ($ k\ped{B} = 1 $ here and in the following). Given the time evolution operator 
	\begin{equation}\label{eq:evolutionoperator}
		U(t, t_{0}) = \timeordering\exp(-\iu \int_{t_{0}}^{t} \ham(\tau) \dd{\tau}),
	\end{equation}
	where $ \timeordering $ is the time-ordering operator, then the density operator at any time $ t $ can be calculated as
	\beq\label{eq:densitymatrix}
		\rho(t) = U(t, t_{0}) \rho(t_{0}) U^{\dagger}(t, t_{0}).
	\eneq 
	Eventually, the density operator of the reduced system at time $ t $ is readily found by tracing out the bath degrees of freedom,
	\beq\label{eq:traceoverboson}
		\rho_{S}(t)= \tr_{B}{\rho(t)},
	\eneq
	thus allowing for the evaluation of any observable of the reduced system. 
	
	A useful numerical approach to calculate $ U(t, t_0) $ is the short-iterative Lanczos method (SIL), which can be employed to propagate the full system-bath quantum state $\ket{\Psi(t)}$ at time $t$, once the starting state $\ket{\Psi(t_{0})}$ is known. This technique combines a projection scheme of the full Hamiltonian~\eqref{eq:system-bath-hamiltonian} to a reduced state space and exact diagonalization methods.
	While conventional approaches describe the influence of the bath degrees of freedom on the reduced systems in terms of an analytically exact effective interaction potential, here the main difficulty resides in finding a suitable truncation scheme of the bath Hilbert space, which could successfully describe the dynamics of $ \rho_S(t) $, at least in a range of model parameters.         
	
	To pursue this goal, we start by discretizing the bosonic spectrum by considering $ M $ equally spaced modes, having frequencies
	\begin{equation}\label{eq:bosonic-modes}
		\omega_k = \frac{\omega\ped{c}}{M} \, k, \qquad k = 1, \dots, M.
	\end{equation}
	The bath space state is spanned by the basis $ \Set{\ket{n_{1},n_{2},\dots,n_{M}}} $, where $ n_{k} = 0, \dots, N\ped{max} $ is the number of excitations in mode $ k $, up to a cut-off $ N\ped{max} $. We integrate Eq.~\eqref{eq:spectral-function-continuous} around each mode, and extract the couplings $ g_k $ which are able to reproduce the correct spectral density of the bath up to some desired level of accuracy, controlled by $ M $. For sufficiently large $ M $, the integral can be approximated by the mean value theorem as
	\begin{equation}\label{eq:bosonic-couplings}
		g_k^2 \approx \eta \frac{\omega_k^s}{\omega\ped{c}^{s-1}} \var{\omega} \equiv \eta \omega\ped{c}^2 \frac{k^s}{M^{s+1}}, 
	\end{equation}
	where $ \var{\omega} = \omega\ped{c} / M $. This uniform sampling is the simplest choice, and allows us to reach convergence in all the investigated regimes, as we will show in the next section. Different samplings have also been proposed in the literature~\cite{de-vega:discretize-baths,bulla:nrg1,bulla:nrg2,zhang:sub-ohm-sbm}.
	
	Further, the truncation scheme to be performed on the set of bath states clearly depends on the value of the coupling strength $ \eta $. As evident from Eq.~\eqref{eq:interaction-hamiltonian}, the creation or annihilation of a boson in a certain state $k$ leads to a variation in the occupation number $n_{k}$ with respect to its thermal equilibrium value $n_{k}\api{eq}$, fixed by the Boltzmann distribution. In the following, we will denote as $ N\ped{ph} $ the absolute maximum number of bosonic excitations, with respect to the thermal equilibrium. Performing the truncation of the Hilbert space to those states with $ \DELTA n_{k}=n_{k}-n_{k}\api{eq} = \Set{0, \pm 1, \pm 2,\dots, \pm N\ped{ph}}$, with $ \sum_k \abs{\DELTA n_k} \leq N\ped{ph} $, an exact description of the system-bath dynamics can be obtained up to terms proportional to $\eta^{N\ped{ph}}$. In the weak coupling regime (WC), we find that a correct description can be obtained by choosing $ N\ped{ph}=1 $; we emphasize that, at WC, our approach recovers the Lindblad results in the limit of extremely weak coupling strengths. For increasing values of $ \eta $, we can fine-tune our results by progressively adding more states to the bath Hilbert space, corresponding to multiple excitations from the equilibrium state; the computational resources needed to simulate the system dynamics at these couplings are necessarily heavier, but calculations remain affordable in the intermediate coupling (IC) regime, where $ N\ped{ph} = 3 $ is enough to get a good quantitative description of the dynamics. As a consequence, this approach is well-suited to describe the correct physical behavior of the system in a parameter range going from weak to intermediate coupling.

	
	

	
	Once the final set of basis states has been fixed, an iterative calculation of the state $\ket{\Psi(\tf)}$ can be set up for any final time $ \tf $ in the following way. First, we divide the entire time interval in subintervals of fixed duration $\dd{t}$. Then, for every fixed time interval $\rng{t}{t+\dd{t}}$, we evaluate the Hamiltonian at midpoint and project it onto the subspace $\mathcal{K}=\Set{\ket{\Psi(t)}, \ham\ket{\Psi(t)}, \dots, \ham^{n}\ket{\Psi(t)}}$, where $\ket{\Psi(t)}$ is the full system state at time $ t $ and $n$ is the minimum number of vectors needed to achieve convergence. An orthonormal basis of vectors in $\mathcal{K}$ is given by the set of Krylov vectors $\Set{\ket{\Phi_{k}}}_{k=1}^{n}$, obtained by recursive Gram-Schmidt orthogonalization techniques. The reduced Hamiltonian $\tilde{\ham}$ in the $ n $-dimensional Krylov subspace can thus be obtained as
	\beq
	\tilde{\ham}=P \ham P^{\dagger},
	\eneq
	where $P$ is the projector operator into the Krylov subspace at time $t$; following the chosen time discretization, the evolution operator $\tilde{U}(t+\dd{t},t)$ can be recast as follows:
	\beq\label{eq:reducedevolution}
	\tilde{U}(t+\dd{t},t)\simeq \exp[-\iu \tilde{\ham}(t+\dd{t}/2) \dd{t}].
	\eneq
	
	The minimum dimension $n$ to achieve convergence depends on $ \dd{t} $; its typical values are of the order of \numrange{20}{100}, thus allowing the numerical evaluation of Eq.~\eqref{eq:reducedevolution} by means of direct diagonalization of the matrix $\tilde{\ham}\qty(t+\dd{t}/2)$. Eventually, expanding the state $\ket{\Psi(t)}$ in terms of the eigenvectors of $\tilde{\ham}\qty(t+\dd{t}/2)$, the full state of the system at time $t+\dd{t}$ can be evaluated by straightforward matrix products. This procedure turns out to be particularly useful if the matrix $\ham$ is Hermitian, because in that case the reduced matrix $\tilde{\ham}\qty(t+\dd{t}/2)$ has tridiagonal form and thus can be easily diagonalized.
	
	One intrinsic limitation of this approach is that it is valid only up to a specific upper time scale.  
	The minimum frequency $ \omega_1 $ determines the Poincar\'{e} recurrence time $ t\ped{p} = 2\uppi / \omega_1 $, which is an upper limit for the total evolution time that can be studied with this method. After $ t\ped{p} $, the collection of harmonic oscillators ceases to be a good approximation of an ergodic thermal bath. Hence, realistically, this numerical approach is not feasible to study very long time (adiabatic) dynamics, except in the WC regime, where the Hilbert space scales linearly with $ M $, allowing to simulate a large number of modes (up to $ 10^6 $) and, consequently, moderately long times.
	
	On the other hand, short time dynamics is well-reproduced even with a limited number of modes, both in WC and IC. That is where our method proves its usefulness. This allows us to study memory effects, which are considered of great interest in real, experimentally controllable, baths~\cite{CialdiExp,NatureChina}. Nonetheless, our method provides the whole $ \text{system}+\text{bath} $ wave function. With a change of perspective, this could be useful to test the influence of the reduced system over the environment, and this is potentially interesting for studying structured environments with a limited number of degrees of freedom. 	
	
	\section{Qubit models}\label{sec:qubits}
		
		In this section, we will discuss theoretically a number of qubit models that we will study with the SIL method presented in Sec.~\ref{sec:numerics}. We start by quickly reviewing the exactly solvable model known in the literature as the pure decoherence model, which we are going to use as a benchmark to test the accuracy of our numerical algorithm. Then, we will move to the more general spin-boson model in presence of a non-zero transverse field, and finally we will apply SIL to a selected time-dependent case, which is relevant for AQC.
	
	\subsection{Pure decoherence}\label{subsec:qubit-exact}
	
	When there is no tunneling in the Hamiltonian~\eqref{eq:tls-hamiltonian}, \ie, $ H_S = -\epsilon(t)\sigma_z $, the dynamics of the two qubit states are decoupled. The full Hamiltonian~\eqref{eq:system-bath-hamiltonian} can be diagonalized using the Lang-Firsov unitary transformation $ \mathcal{U} = \exp(-\mathcal{S}) $, where
	\begin{equation}\label{eq:lang-firsov}
		\mathcal{S} = \sigma_z \sum_k \frac{g_k}{\omega_k} \qty(b_k^\dagger - b_k),
	\end{equation}
	which shifts the center of each harmonic oscillator according to the qubit state~\cite{mahan:many-particle}. The transformed Hamiltonian is diagonal and its eigenstates are coherent states in the bath degrees of freedom. Notice that the Hamiltonian commutes with $ \sigma_z $, hence the occupations of the TLS states are conserved quantities.
	
	We consider $ \epsilon(t) = \epsilon $ as a fixed energy scale so that the reduced spectral gap $ \Delta $ is constant in time and equal to $ 2\epsilon $, and prepare the state at $ t = 0 $ as an uncorrelated product of the Boltzmann equilibrium state at inverse temperature $ \beta $ and the positive eigenstate $ \ket{\hat{x}; +} $ of the operator $ \sigma_x $. Notice that an analytical solution exists for this trivial case even if $ \epsilon $ is time-dependent; the features of the solution are qualitatively similar to the time-independent case, and so are the predictions of the SIL method, hence we focus here on the time-independent case for simplicity. We measure the instantaneous properties of the reduced system at any time during the dynamics, in particular we study the mean values $ \ev{\sigma_x(t)} $ and $ \ev{\sigma_z(t)} $, which, for the pure decoherence model, are related to decoherence and thermal relaxation, respectively. As $ \Gamma(t) = 0 $, the reduced system does not relax ($ \ev{\sigma_z(t)} = \ev{\sigma_z(0)} = 0 $) and the only non-trivial quantity is $ \ev{\sigma_x(t)} $, which can be evaluated analytically provided that the initial state is factorized~\cite{mahan:many-particle}:
	\begin{equation}\label{eq:exact-sigmax-mean}
		\ev{\sigma_x(t)} = \cos(2\epsilon t) \exp[-\eta K(t, \beta)],
	\end{equation}
	where we introduced the decoherence function
	\begin{equation}\label{eq:exact-decoherence}
		K(t, \beta) \equiv \frac{8}{\eta}\sum_k \frac{g_k^2}{\omega_k^2}\sin[2](\frac{\omega_k t}{2}) \coth(\frac{\beta\omega_k}{2}).
	\end{equation}
	Notice that this function, as defined, is coupling-independent in the continuous limit.
	
	The same model can be solved using the Lindblad equation for the reduced density matrix (see App.~\ref{app:lindblad}). We do not enter the details of the calculation here, and report the result~\cite{albash:decoherence}:
	\begin{equation}\label{eq:exact-sigmax-lindblad}
		\ev{\sigma_x(t)}\ped{L} = \cos(2\epsilon t) \exp[-2\gamma(0) t],
	\end{equation}
	where $ \gamma(\omega) $ is defined in Eq.~\eqref{eq:spectral-functions-lindblad}.
	
	Comparing Eqs.~\eqref{eq:exact-sigmax-mean} and~\eqref{eq:exact-sigmax-lindblad}, we see that the coherent part of the mean value (the cosine function) is well-predicted by the Lindblad theory; however, the Lindblad decoherence function recovers only the adiabatic limit $ t\to\infty $ (with $ \beta < \infty $) of the actual decoherence function, as the following limit holds:
	\begin{equation}\label{eq:limit-delta}
		\lim\limits_{t\to\infty}\frac{\eta K(t,\beta)}{t} = 2 \gamma(0).
	\end{equation}
	Notice that Eq.~\eqref{eq:exact-sigmax-lindblad} always fails to predict the correct behavior of the solution at $ \beta\to\infty $. 
	The reason of this discrepancy is that, below $ \tau_B = \beta / \uppi $, which is the characteristic decay time of the self-correlation function of the bath $ \mathcal{B}(t) =\ev{B(t) B(0)} $, the finite-temperature contribution to the decoherence function, $ \upDelta K(t,\beta) =  K(t,\beta) - K(t,\infty) $, is negligibly small. For $ t \gg \tau_B $, the relevant term is $ \upDelta K(t,\beta) $, as it grows as a power law, while $ K(t,\infty) $ grows logarithmically~\cite{Zanardi}. The Lindblad approximation always disregards $ K(t,\infty) $, meaning that it always fails at small times with respect to $ \tau_B $. On the other hand, $ \tau_B $ diverges at low temperatures, hence the Lindblad approximation is inadequate in this limit. 
	A numerical analysis on this point is proposed in App.~\ref{app:decoherence}.

	We simulated the same system at $ T = 0 $ using our numerical SIL method, following the scheme depicted in Sec.~\ref{sec:numerics}, with a cut-off frequency $ \omega\ped{c} = 10\epsilon $ and a collection of $ M = 200 $ modes. The Poincar\'{e} recurrence time is $ \epsilon t\ped{p} = 40\uppi $; we purposely restricted the dynamics up to the shorter time $ 40 / \epsilon $ to limit spurious effects arising because of recurrence. We set the number of Lanczos iterations at each time step to be $ 30 $. The reduction scheme concerning relevant phonon processes has been tested by considering $ N\ped{ph} = \Set{1,2,3} $ excitations from the bosonic vacuum, \ie, the thermodynamic equilibrium state at zero temperature. This led to Hilbert spaces of the full system of dimensions $ D_1 = \num{402} $, $ D_2 = \num{40602} $, $ D_3 = \num{2747402} $, respectively. We conducted our analysis for the two representative couplings $ \eta = 10^{-4} $ (WC) and $ \eta = 10^{-2} $ (IC). In the following plots, we present our data about the instantaneous relative error 
	\begin{equation}\label{eq:relative-error}
		\delta(t) \equiv \frac{\ev{\sigma_x(t)}\ped{SIL} - \ev{\sigma_x(t)}\ped{th}}{\ev{\sigma_x(t)}\ped{th}}
	\end{equation}
	of the simulated solution $ \ev{\sigma_x(t)}\ped{SIL} $ with respect to the analytical theoretical solution $ \ev{\sigma_x(t)}\ped{th} $ (Eq.~\eqref{eq:exact-sigmax-mean}).
	
\begin{figure}[t]
		\centering
		\includegraphics[width=\linewidth]{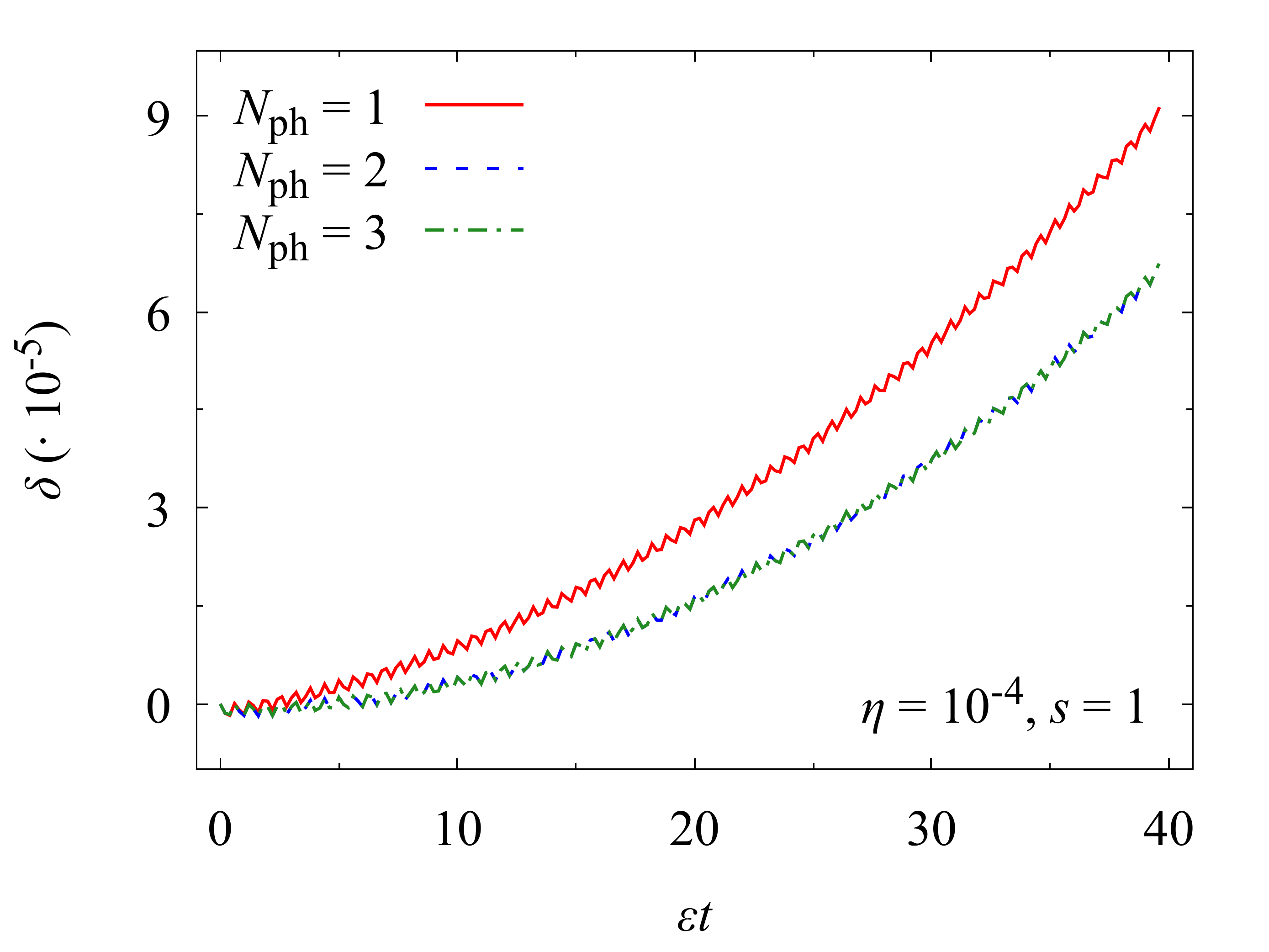}
		\caption{Relative error in the SIL approximation of $ \ev{\sigma_x(t)} $ with respect to the analytical solution, for an Ohmic bath ($ s = 1 $) at $ T = 0 $, coupled with a constant $ \eta = 10^{-4} $. }
		\label{fig:exact200modierroreohmiceta00001nph123}
	\end{figure}

\begin{figure}[t]
		\centering
		\includegraphics[width=\linewidth]{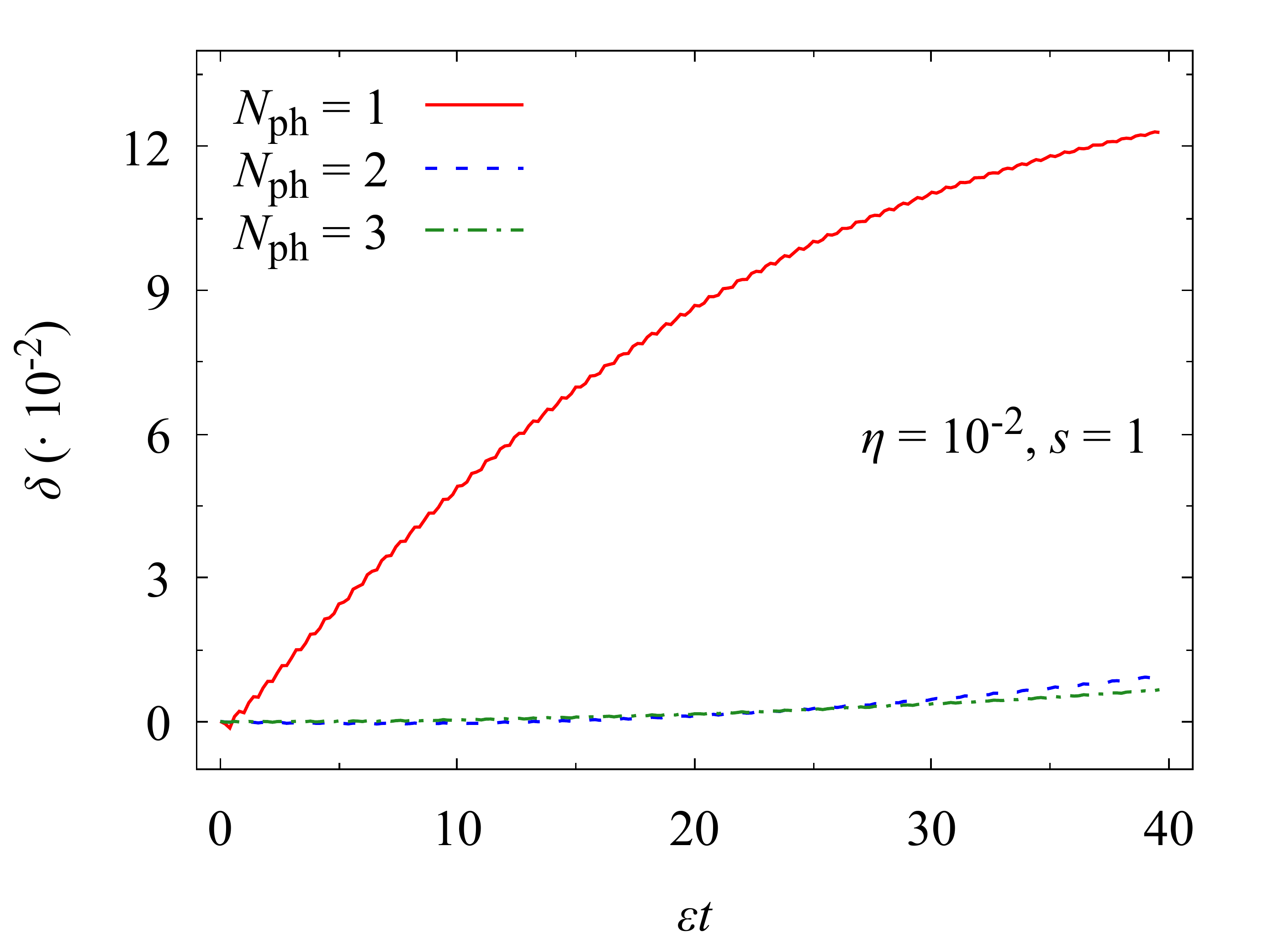}
		\caption{Relative error in the SIL approximation of $ \ev{\sigma_x(t)} $ with respect to the analytical solution, for an Ohmic bath ($ s = 1 $) at $ T = 0 $, coupled with a constant $ \eta = 10^{-2} $. }
		\label{fig:exact200modierroreohmiceta001nph123}
	\end{figure}

	Fig.~\ref{fig:exact200modierroreohmiceta00001nph123} shows the relative error of the approximation for an Ohmic bath in the WC regime. Including a single bosonic excitation per mode reproduces quantitatively the exact solution with a relative error of $ 10^{-5} $. The result is almost unchanged when multiple-phonon processes are included, and the gain in accuracy saturates when $ N\ped{ph} = 2 $, indicating that the increasing trend in the curves is only related to the discreteness of our bath, rather than to the processes cut-off, and can thus be improved by including more modes. As expected, this scenario changes in the IC regime. In fact, as evident from Fig.~\ref{fig:exact200modierroreohmiceta001nph123}, multiple-phonon processes play an important role, although $ N\ped{ph} = 3 $ still provides a good approximated solution. An analogous discussion for sub-Ohmic and super-Ohmic dissipations is proposed in App.~\ref{app:errors-sub-super}. 
	
\subsection{Spin-boson model}\label{subsec:spin-boson}

If both terms in the Hamiltonian of Eq.~\eqref{eq:tls-hamiltonian} are non-zero and time-independent, the model~\eqref{eq:system-bath-hamiltonian} reduces to the well-known SBM, \ie, $ H_S = -\Gamma \sigma_x - \epsilon \sigma_z $. Although no analytical solution is known for the non-trivial dynamics of $ \ev{\sigma_z(t)} $ and $ \ev{\sigma_x(t)}$, several approximation schemes have been proposed over the last four decades in order to tackle this problem.

In the traditional formulation of the problem~\cite{Leggett}, the initial conditions are set such that, at $ t = 0 $, the qubit is prepared into an eigenstate $ \ket{\hat{z}; +} $ of the operator $ \sigma_{z} $. The initial state of the qubit-bath system is factorized, the bath is at thermal equilibrium at temperature $ T = 1/\beta $, and the interaction of the qubit with its surroundings is modeled by the Hamiltonian~\eqref{eq:interaction-hamiltonian}. The solution of the problem consists in finding an approximation for the reduced density matrix of the qubit $ \rho_{S}(t) $ at time $t$.

The Lindblad equation~\eqref{eq:lindblad-equation}, based on Born and Markov approximations, provides a closed-form solution for the density matrix $ \rho_{S}(t) $, as shown in Eq.~\eqref{eq:densitylindsbm}. 
However, while representing a useful tool to reduce the complexity of the problem, the Lindblad equation suffers from several limitations, \eg, it is expected to be valid only in the weak coupling limit and if non-Markovian effects can be safely neglected.    


A noticeably broader approach, based on a standard path-integral formulation~\cite{FEYNMAN63}, allows to perform the sum over all bath degrees of freedom in an influence functional~\cite{Leggett,Grifoni}, affecting the dynamics of $ \rho_S (t) $. The resulting expression for $ \rho_S (t) $ is analytically intractable, and can be attacked using different approximation schemes, among which are the NIBA~\cite{Leggett} and its IC extension, WIBA~\cite{Grifoni}; numerically exact methods have also been thoroughly explored~\cite{LeHurNRG1}.

In what follows, we analyze the dynamics of SBM by means of the SIL technique described in Sec.~\ref{sec:numerics}. We first restrict to the unbiased case $ \epsilon = 0 $, \ie, $\ham_S=-\Gamma \sigma_x $, and, by strict analogy with Sec.~\ref{subsec:qubit-exact}, we discuss the dynamics in the limit $ T = 0 $. We choose a cut-off frequency $\omega\ped{c}=10\Gamma$, take the coupling parameter $\eta$ in the range $\eta = \numrange[range-phrase = \text{~to~},exponent-product=\cdot]{5e-4}{1e-1}$, and assume $s=\Set{1/2,1,2}$; in addition, following this choice of parameters, we perform the basis truncation including up to three excitations per mode ($N\ped{ph}=3$).              
We prepare the system at time $t=0$ in a linear combination of the basis states at fixed starting values $\ev{\sigma_x(0)} = \ev{\sigma_z(0)} = 1/2 $, \ie, $ \ket{\psi(0)} = \cos(\xi/2) \ket{\hat{z}; +} + \sin(\xi/2) \exp(\iu \phi) \ket{\hat{z}; -} $, with $ \xi = \uppi / 3$  and $ \phi = \acos\,\!(1/\sqrt{3}) $.
Then, we calculate the time-evolved mean values $ \ev{\sigma_x(t)} $ and $\ev{\sigma_z(t)}$, extracted from the reduced density matrix $\rho_{S}(t)$, and eventually compare them with their analytical closed-form counterparts obtained from the Lindblad equation.

\begin{figure}[t]
	\centering
	\includegraphics[width=\linewidth]{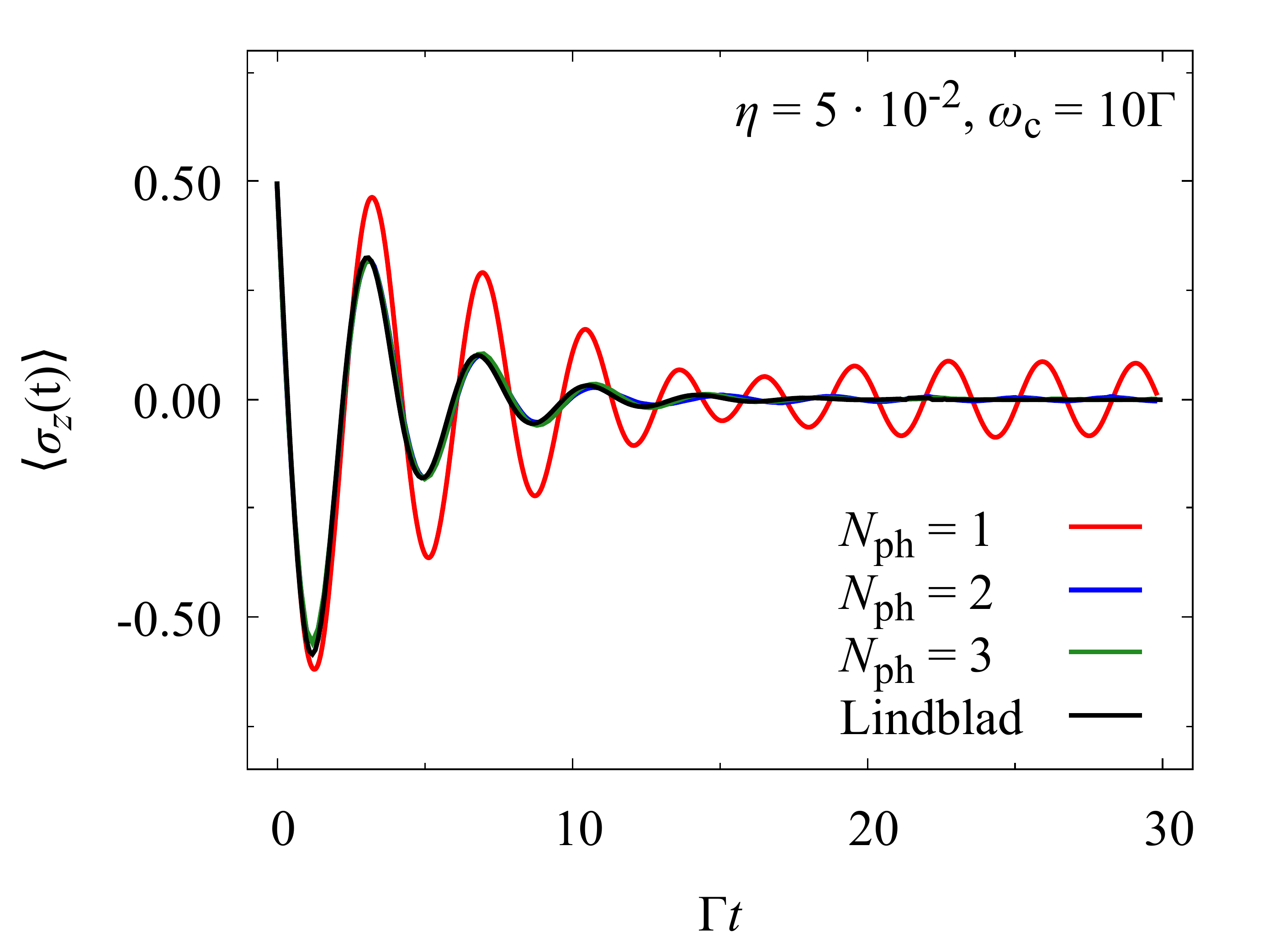}
	\caption{Time evolution of $\ev{\sigma_z(t)}$ for an unbiased qubit in an Ohmic bath ($ s = 1 $), having chosen $ \eta = 5\cdot 10^{-2} $, $\omega\ped{c}=10\Gamma$ and $T=0$. We fixed $M=\Set{1000,500,300}$, for $N\ped{ph} = \Set{1, 2 , 3}$, respectively. SIL results are plotted against the Lindblad curve (solid black curve), from Eq.~\eqref{eq:Lindspinboson}.}
	\label{fig:sigmaz_t_ohmic}
\end{figure}

In Fig.~\ref{fig:sigmaz_t_ohmic}, we show the results for $ \ev{\sigma_z(t)} $ in the IC regime, in the case of Ohmic dissipation, for different values of the maximum number of excitations per mode $N\ped{ph}$, compared with the result predicted by the Lindblad equation in Eq.~\eqref{eq:Lindspinboson}. Choosing a minimum value of $N\ped{ph}=2$, the time evolution of $ \ev{\sigma_z(t)} $ converges to the exact physical behavior, which shows underdamped oscillations due to decoherence effects. It follows that at long times the equilibrium value $\sigma_z\api{eq}=0$ is reached and the system completely loses its coherence.

\begin{figure}[t]
	\centering
	\includegraphics[width=\linewidth]{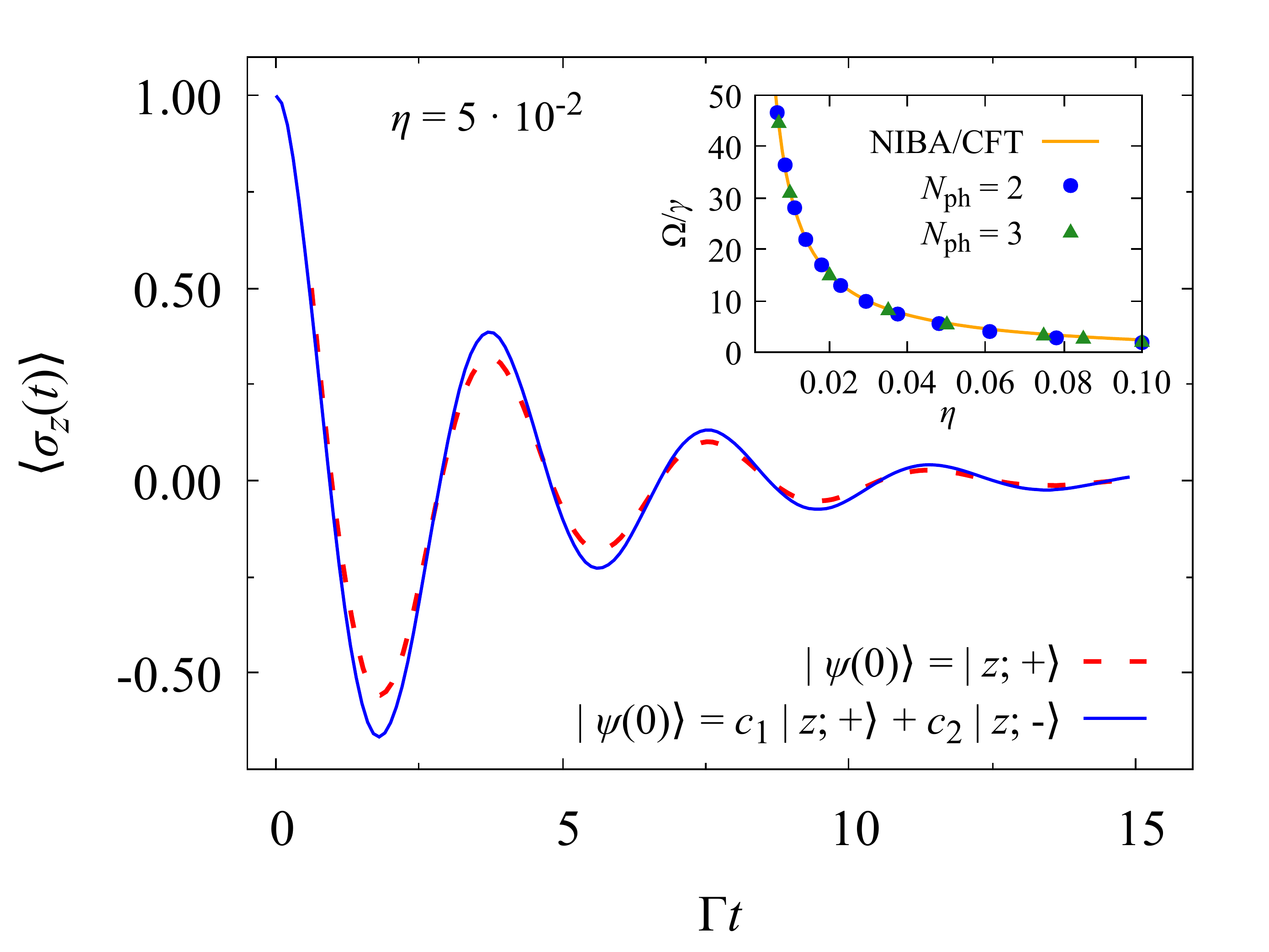}
	\caption{Time evolution of $\ev{\sigma_z(t)}$ for an unbiased qubit in an Ohmic bath ($ s = 1 $), with $ \eta = 5\cdot 10^{-2} $, $\omega\ped{c}=10\Gamma$ and $T=0$, and two different starting conditions: the traditional $ \ket{\hat{z}; +} $ and the state $ c_1 \ket{\hat{z}; +} + c_2 \ket{\hat{z}; -} $, with $ c_1 = \cos\xi  $ and $ c_2 = \exp(\iu\phi)\sin\xi $ (see the main text for their definition). In the inset, SIL results for the quality factor as a function of the coupling parameter, compared with conformal field theory and NIBA (solid orange line)~\cite{orth-lehur}; in the IC regime, increasing the phonon number is necessary to improve the accuracy.}
	\label{fig:sigmaz-starting-condition-renormalized-gap}
\end{figure}

Notice that, as expected, these features do not depend on the starting condition. In the main plot of Fig.~\ref{fig:sigmaz-starting-condition-renormalized-gap}, we support this statement by comparing the dynamics for our choice of the initial state with the more traditional $ \ket{\psi(0)} = \ket{\hat{z}; +} $. Here is clearly seen that both the decay rate and the oscillation frequency are preserved; in order to emphasize this, we shifted one of the curves to make them in phase. Simulation parameters are the same as Fig.~\ref{fig:sigmaz_t_ohmic} and we chose $ N\ped{ph} = 2 $. The functional form of $ \ev{\sigma_z(t)} $ is $ A \cos(\Omega t + \mu) \exp(-\gamma t) $, where the frequency $\Omega$ and the damping factor $\gamma$ are related to the tunneling amplitude. It is known by theoretical arguments~\cite{Leggett} that the interaction with the environment is responsible for a renormalization of the tunneling amplitude (and, correspondingly, of the spectral gap of the qubit system), depending on the coupling strength, of the form 
\begin{equation}\label{eq:renormalized-gap-nrg}
	\Gamma\ped{r}\api{(RG)} = \Gamma \qty(\frac{2\Gamma}{\omega\ped{c}})^{\frac{2\eta}{1-2\eta}}.
\end{equation}
Despite its limitations, NIBA yields correct predictions for the quality factor $\Omega/\gamma$ of the damped oscillations, in agreement with conformal field theory~\cite{orth-lehur}. It reads   
\begin{equation}\label{eq:renormalized-gap-niba}
	\frac{\Omega}{\gamma} = \cot\frac{2 \pi \eta}{2(1-2\eta)}.
\end{equation}
In the inset of Fig.~\ref{fig:sigmaz-starting-condition-renormalized-gap}, we show that the SIL method succesfully recovers the behavior of the quality factor in the entire range of investigated coupling strengths. 

\begin{figure}[tb]
 	\centering
 	\includegraphics[width=\linewidth]{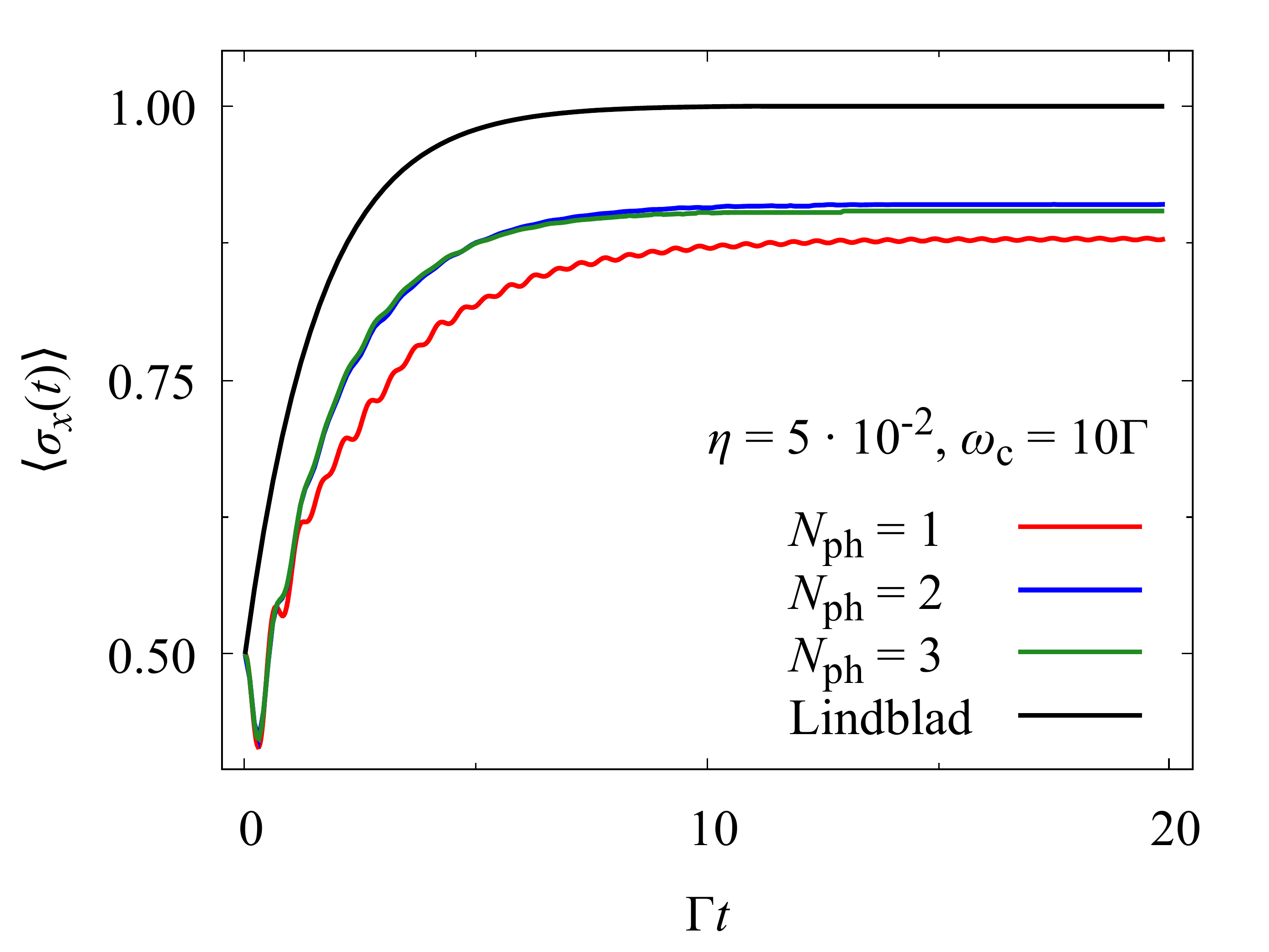}
 	\caption{Time evolution of $\ev{\sigma_x(t)}$ for the same parameter values as in Fig.~\ref{fig:sigmaz_t_ohmic}. The non-monotonic region occurs at short times $0 <\Gamma t<2.5 $ while at long times the curve saturates to a well-definite equilibrium value.}
 	\label{fig:sigmax_t_ohmic}
 \end{figure}

\begin{figure}[tb]
 	\centering
 	\includegraphics[width=\linewidth]{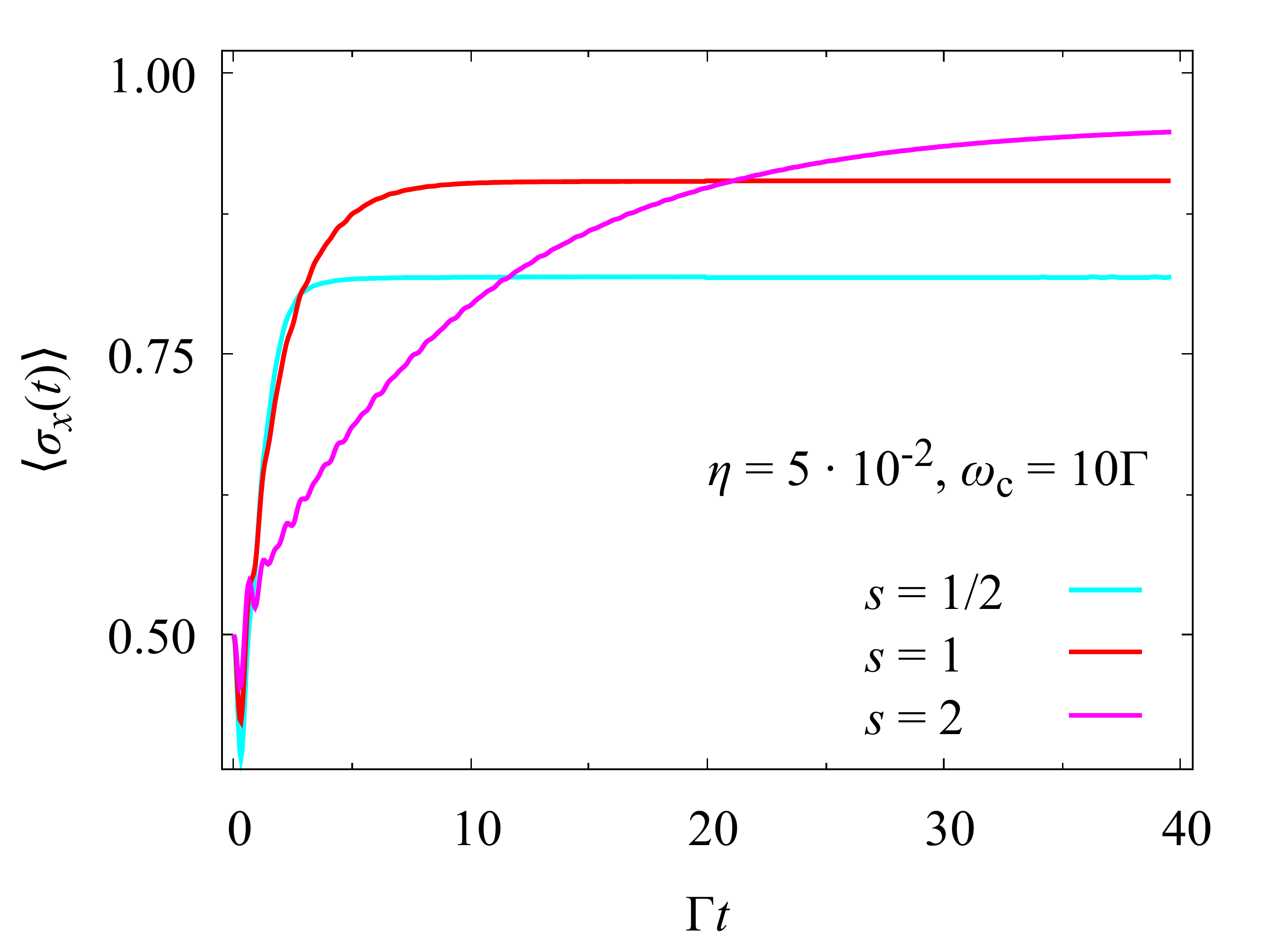}
 	\caption{Time evolution of $\ev{\sigma_x(t)}$ for the same parameter values as in Fig.~\ref{fig:sigmaz_t_ohmic}, for three different dissipations considered $s=\Set{1/2,1,2}$. Both the equilibration times and saturation values depend on $s$.}
 	\label{fig:sigmax_varis}
 \end{figure}

On the other hand, in Fig.~\ref{fig:sigmax_t_ohmic} we analyze the numerical results for the time evolution of $ \ev{\sigma_x(t)} $ obtained by means of SIL technique, plotted against the result predicted by the Lindblad equation reported in Eq.~\eqref{eq:Lindspinboson}; our results exhibit a non-monotonic behavior at short times, 
 while a prominent saturation behavior at long times can be observed, for every value of $N\ped{ph}$. Analogous properties hold for the time evolution of this observable in the sub-Ohmic and super-Ohmic cases (see Fig.~\ref{fig:sigmax_varis}), provided that the analysis is restricted to WC and IC regimes. As shown in Fig.~\ref{fig:sigmax_varis}, the three saturation curves, in the same parameter region as in Fig.~\ref{fig:sigmaz_t_ohmic}, clearly differ in the equilibration times as well as in the equilibrium values $\sigma_x\api{eq}$.
 
 As can be inferred from Fig.~\ref{fig:sigmax_t_ohmic}, our results remarkably differ from the Lindblad one, because the latter predicts as the long-time stationary value the one corresponding to the ground state of the qubit Hamiltonian disentangled from the bath. Instead, our calculations show that the stationary value is related to the ground state of the qubit-bath system: at long times, qubit and bath remain entangled, as expected at equilibrium.
 
 While such a striking difference can be observed in the equilibrium values of $\ev{\sigma_x(t)}$ obtained by using SIL and the Lindblad equation, the relaxation rates are very similar in the two approaches. As a deeper analysis of Figs.~\ref{fig:sigmaz_t_ohmic} and~\ref{fig:sigmax_t_ohmic} shows, the Lindblad result for $\ev{\sigma_z(t)}$ qualitatively agrees with the SIL result, correctly predicting the decoherence behavior, which takes place in a time $T_{2}$ depending on the energy gap $ \Delta = 2\Gamma $, temperature and the damping parameter $ \eta $ (see Eqs.~\eqref{eq:Lindspinboson}).     
Note also that the time dependence obtained by the SIL method with $ N\ped{ph} = 1 $ fails to recover the correct physical behavior suggesting that, as expected, the Lindblad solution includes multiple uncorrelated scattering processes. On the other hand, as previously discussed, the Lindblad result for $\ev{\sigma_x(t)}$---equal to the difference in populations of states $ \ket{\hat{x}; \pm} $---saturates towards the wrong asymptotic value after a time $T_{1}=T_{2}/2$ (see App.~\ref{app:spinboslind}). In this case, correlations among multiple scattering processes, correctly included by our approach, play a relevant role. Referring to the diagrammatic theory, our approach includes vertex corrections which are disregarded in the Lindblad approximation. 

\begin{figure}[tb]
	\centering
	\includegraphics[width=\linewidth]{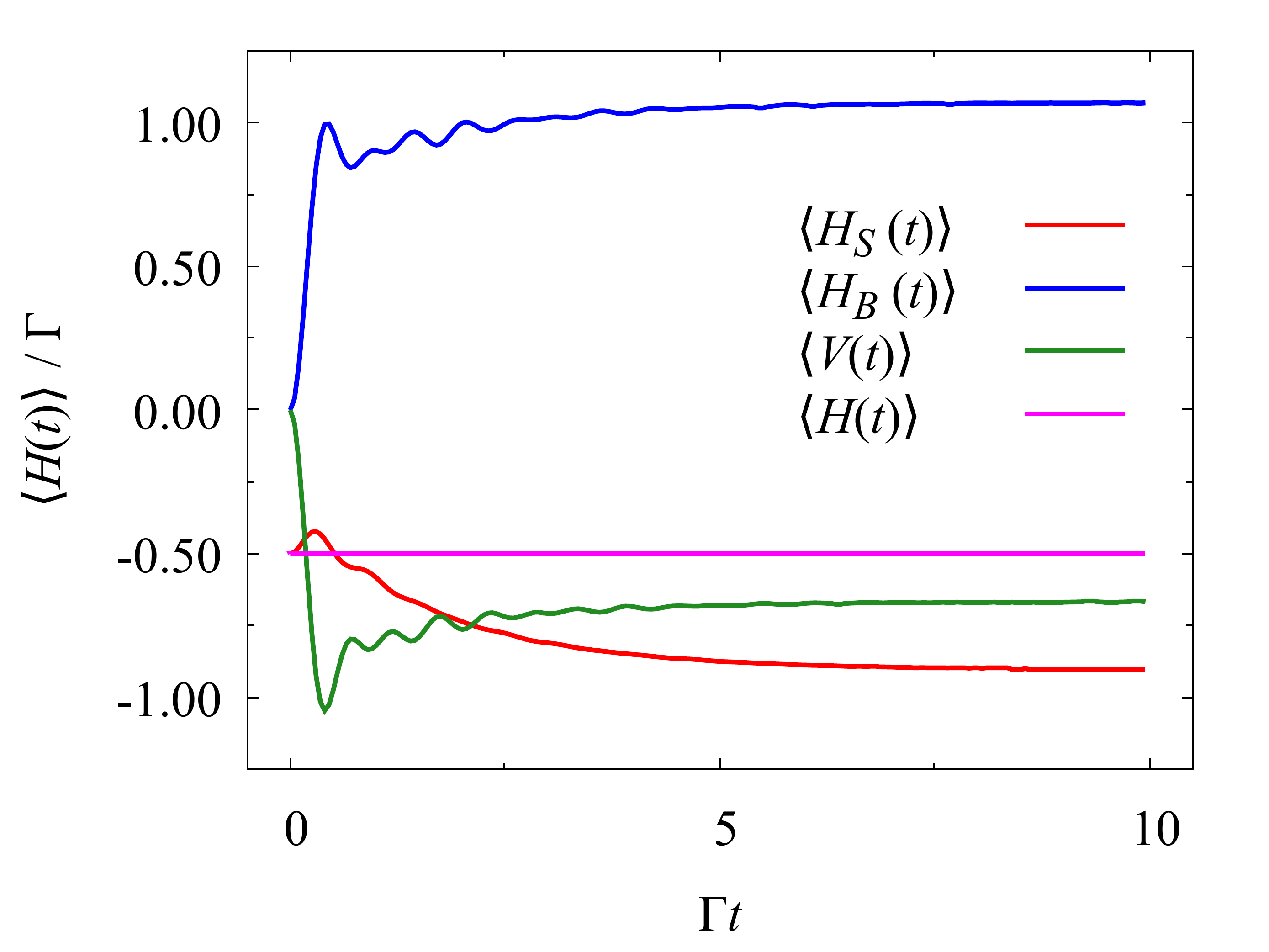}
	\caption{Time evolution of $\ev{\ham_S(t)}$, $\ev{\ham_B(t)}$, $\ev{V(t)}$, and $\ev{\ham(t)}$, in units of $\Gamma$, for an Ohmic bath with $M=300$, $N\ped{ph}=3$, $ \eta = 5\cdot 10^{-2} $ and $\omega\ped{c}=10\Gamma$.}
	\label{fig:energie_medie}
\end{figure}

In addition, while the relaxation times are correctly reproduced, we note that the Lindblad approximation in Eq.~\eqref{eq:Lindspinboson} does not take into account the non-monotonic behavior of $\ev{\sigma_x(t)}$ at very short times, as shown in Fig.~\ref{fig:sigmax_t_ohmic}. 
This behavior can be understood by carrying out a detailed analysis of the time evolution of each contribution to the expectation value of the total Hamiltonian in  Eq.~\eqref{eq:system-bath-hamiltonian}. As shown in Fig.~\ref{fig:energie_medie}, at short times the absolute value of the system-bath interaction energy rapidly grows up to an absolute maximum and, as a consequence, both the reduced system and the bath undergo an excitation from their initial states, while the total energy remains constant in time. After this brief transient time, depending on the chosen initial condition, the expectation value of the reduced system energy $\ev{\ham_S(t)}$, as well as $\ev{\ham_B(t)}$ and $\ev{V(t)}$, saturates towards its equilibrium value.

The previous results suggest that, moving from WC to IC, a physical description of the dynamics of the SBM entirely based on the Lindblad equation can suffer from severe limitations, in agreement with theoretical~\cite{breuer:open-quantum} and experimental findings~\cite{CialdiExp, NatureChina}. On the other hand, the SIL approach can successfully reproduce the correct physical scenario in this parameter region. In order to provide evidence for it, we study the equilibrium values $\sigma_x\api{eq}$ as a function of the coupling parameter $ \eta $ for the three different kinds of dissipation mentioned before, choosing the maximum number of excitations per mode up to $ N\ped{ph}=3 $.      

\begin{figure}[tb]
	\centering
	\includegraphics[width = \linewidth]{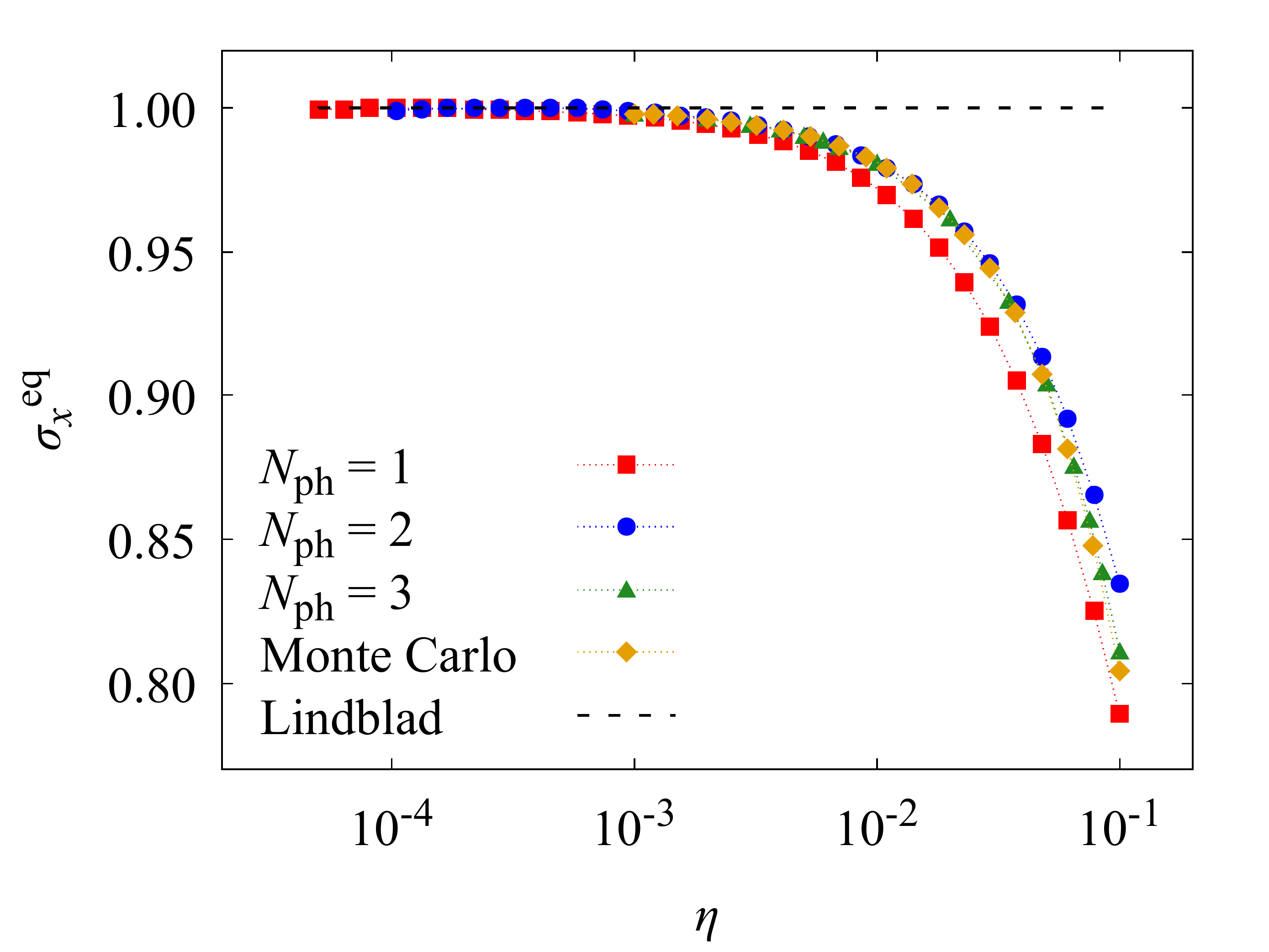}
	\caption{Semi-logarithmic plot of the saturation value $\sigma_x\api{eq}$, extrapolated with exponential fits, as a function of the dimensionless coupling $ \eta $, at $ T = 0 $, for an Ohmic bath ($ s = 1 $). The simulated data of the numerical diagonalization, up to $ N\ped{ph} = 3 $ bosonic excitations from the vacuum state, are compared to Lindblad and Monte Carlo predictions at equilibrium. The ranges of parameters where the physics is ruled either by single or multiple-phonon processes are easily distinguishable by those values of $ \eta $ where the curve at different $ N\ped{ph} $ separate.}
	\label{fig:equilibrium-sx-ohmic}
\end{figure}

\begin{figure}[tb]
	\centering
	\includegraphics[width = \linewidth]{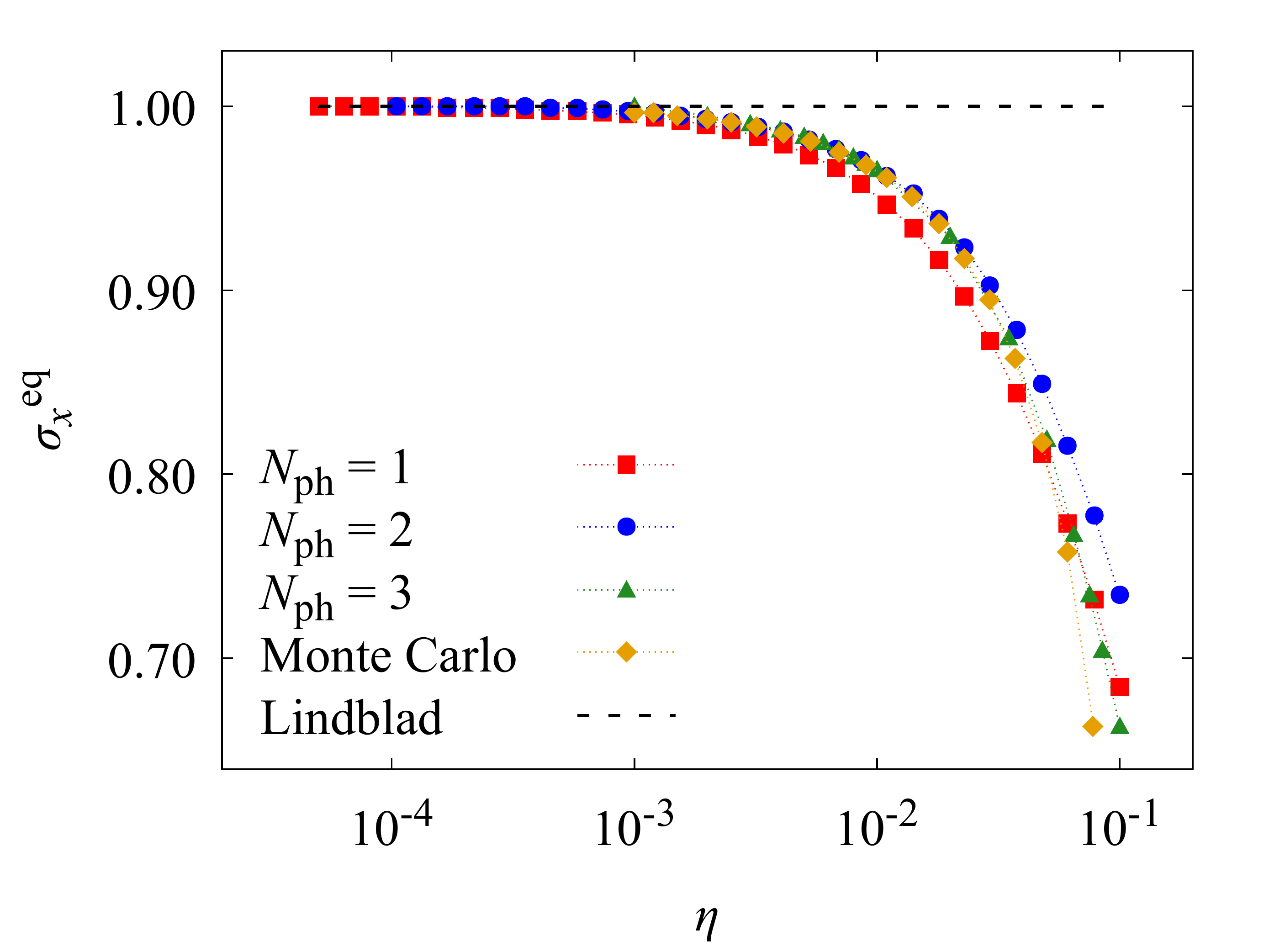}
	\caption{Same as Fig.~\ref{fig:equilibrium-sx-ohmic}, but for a sub-Ohmic bath ($ s = 1/2 $).}
	\label{fig:equilibrium-sx-subohmic}
\end{figure}

\begin{figure}[tb]
	\centering
	\includegraphics[width = \linewidth]{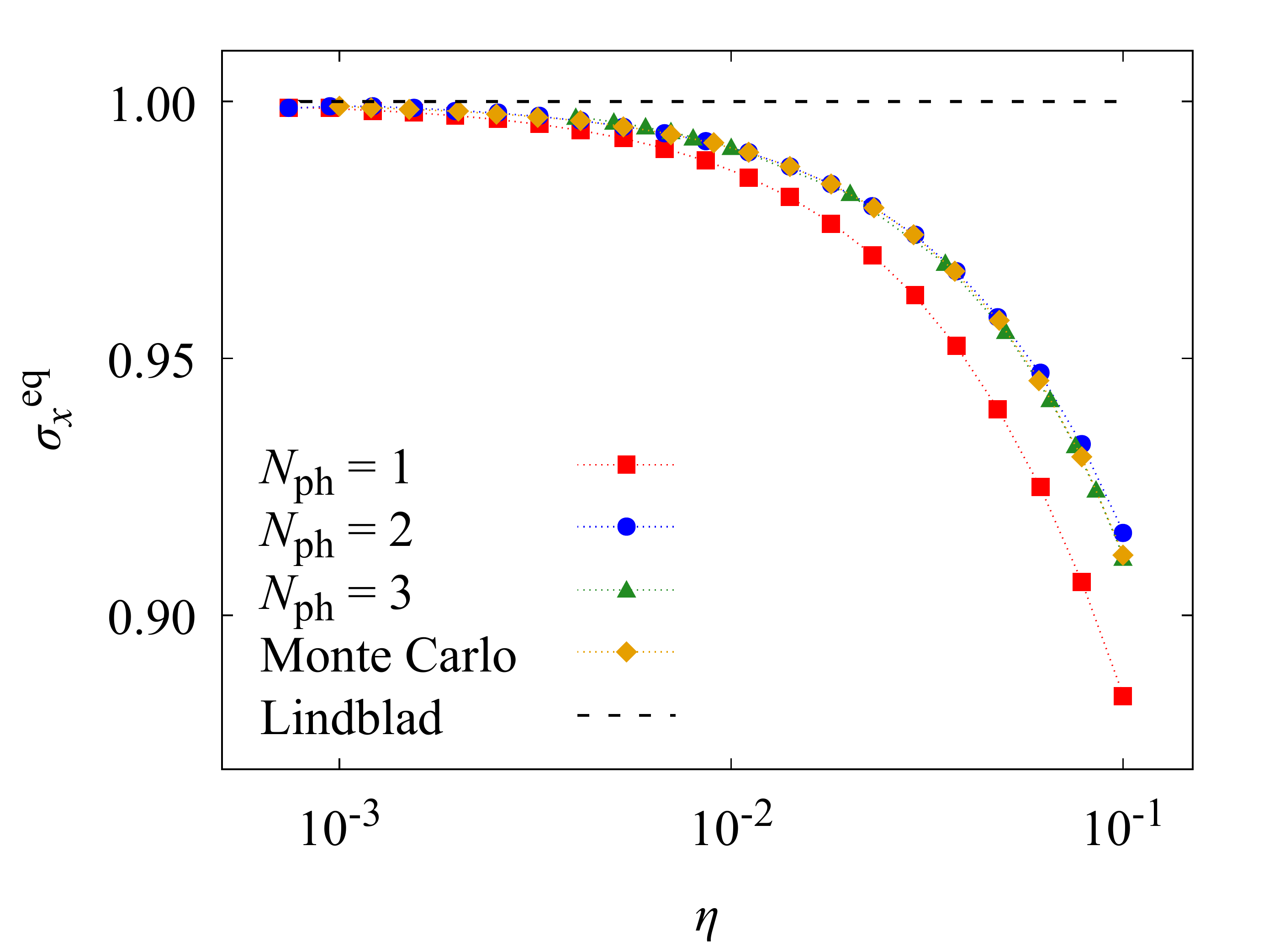}
	\caption{Same as Fig.~\ref{fig:equilibrium-sx-ohmic}, but for a super-Ohmic bath ($ s = 2 $). In presence of a super-Ohmic bath, because of the reduced number of low-energy bosons, equilibrium is reached more slowly and it is difficult to extrapolate $\sigma_x\api{eq}$ at WC.}
	\label{fig:equilibrium-sx-superohmic}
\end{figure}

In order to obtain reliable values of $\sigma_x\api{eq}$, we performed an exponential fit of the numerical results $\ev{\sigma_x(t)}$ and extracted the best estimates of the saturation values. In Figs.~\ref{fig:equilibrium-sx-ohmic},~\ref{fig:equilibrium-sx-subohmic} and~\ref{fig:equilibrium-sx-superohmic}, we show the fitted equilibrium values $ \sigma_x\api{eq} $ as a function of the coupling parameter $\eta$ compared with the Lindblad result. In order to further test the reliability of our calculations, we also plot the equilibrium values calculated using a Monte Carlo approach at thermal equilibrium (orange filled diamonds)~\cite{DeFilippisunpub}. We note that, as the coupling factor becomes larger than $10^{-3}$, the Born-Markov approach misses the correct physical behavior for every bath spectral distribution considered. It follows that, at long times, the unavoidable system-bath entanglement effects start to play a role, noticeably reducing the value of $\sigma_x\api{eq}$. This effect becomes particularly evident in the case of sub-Ohmic dissipation, which shows a rapid decrease of the  $\ev{\sigma_x(t)}$ as $\eta$ reaches $10^{-2}$. This is due to 
the fact that, in this case, the critical coupling strength at which the quantum phase transition of the SBM~\cite{Leggett} occurs is smaller than in the Ohmic case~\cite{bulla:nrg1,bulla:nrg2}, explaining the observed quantitative difference between Monte Carlo data and SIL predictions. On the other hand, in the Ohmic and super-Ohmic case, as far as the coupling factor is weaker than $10^{-1}$, a good physical description can be achieved by truncating the phonon bases to three excitations per mode.

Following these results, we can apply the SIL technique to perform an analogous analysis for the biased case ($\epsilon \neq 0$). Here, the gap between the qubits states changes to a constant value equal to $\Delta=2\sqrt{\epsilon^{2}+\Gamma^{2}}$, the eigenstates being linear superpositions of the computational basis states. The biased case is of particular interest for us, since it has been shown that the NIBA, predicting the qubit localization in the state $ \ket{\hat{z}; -} $ at long times, fails to describe the correct physical behavior~\cite{Leggett,Grifoni}. We can therefore further test the predictions of our numerical technique by analyzing the asymptotic behavior of $\ev{\sigma_z(t)} $. By turning back to the $ \ket{\hat{z}; \pm} $ basis, we prepare the qubit at initial time in the state $ \ket{\hat{z}; +} $ and simulate the time evolution of the biased system in WC, by fixing the values $\epsilon=-\Gamma$, $\eta= 5\cdot 10^{-3} $ and $ T=\Set{0, 0.1} $ in units $ \Gamma $.  

\begin{figure}[tb]
	\centering
	\includegraphics[width=\linewidth]{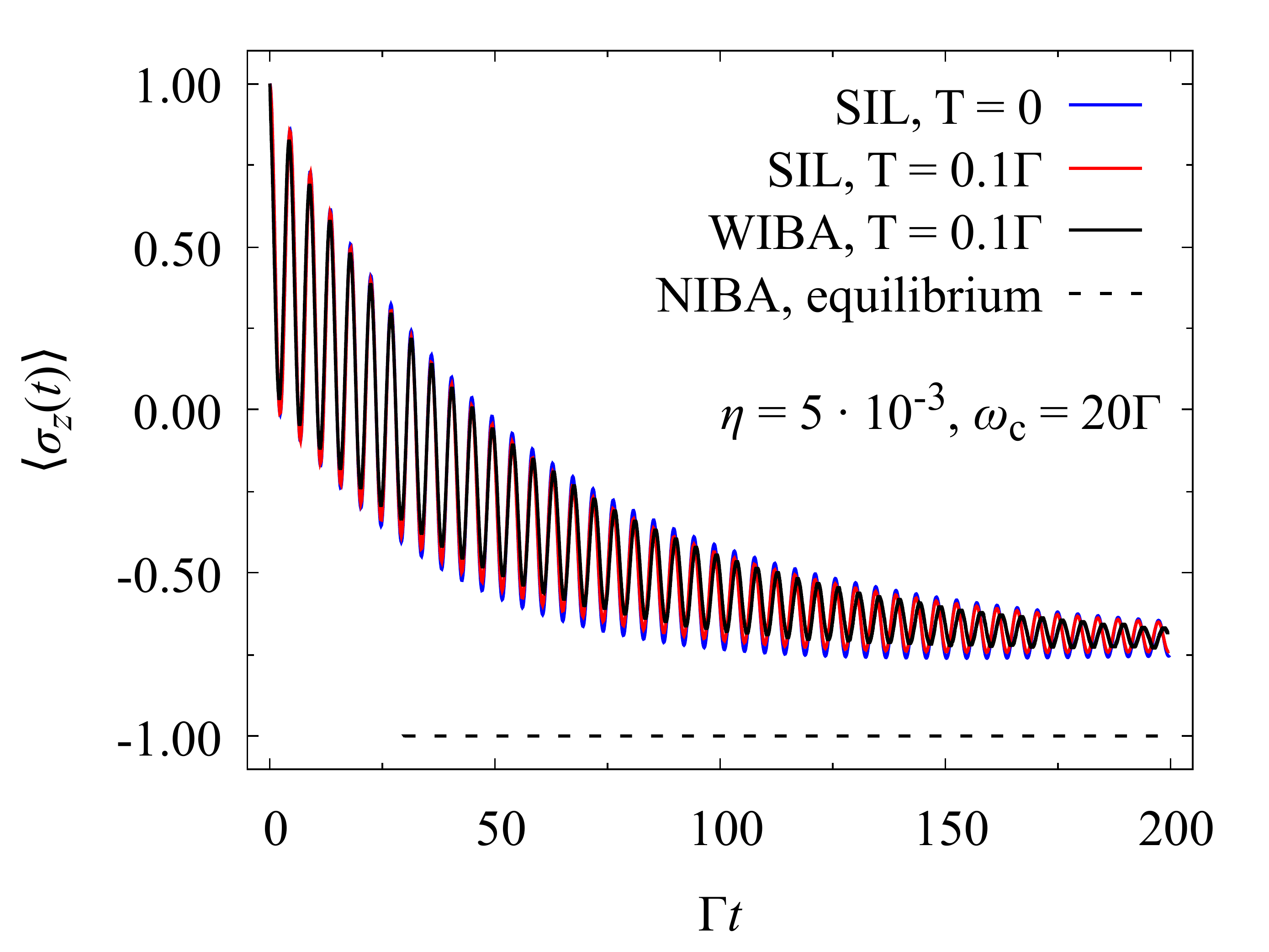}
	\caption{Time evolution of $ \ev{\sigma_z(t)}$ for the biased case and Ohmic dissipation, obtained by choosing $\epsilon=-\Gamma$ , $\eta= 5 \cdot 10^{-3} $, $ \omega\ped{c}=20\Gamma$, $ N\ped{ph} = 2 $, and $T=\Set{0, 0.1}$ in units $ \Gamma $ (red and blue solid curves). The asymptotic value, while differing from that predicted by NIBA, fairly agrees with WIBA results (solid black curve), extracted from Ref.~\cite{Grifoni}.}
	\label{fig:biased_sigmaz_t} 
\end{figure}

As shown in Fig.~\ref{fig:biased_sigmaz_t}, the numerical results for $ \ev{\sigma_z(t)} $ clearly indicate an asymptotic value that, while differing from the NIBA ($ \ev{\sigma_z(\infty)}\ped{NIBA}=\tanh(\beta \epsilon/2)$)~\cite{Leggett}, is consistent with that obtained by means of WIBA approach~\cite{Grifoni}. It follows that our method can provide an accurate description of correlation effects, and it can be fruitfully used to describe the physics of these system even in the IC regime. In addition, a more detailed numerical analysis may be pursued in order to measure in a systematic way the differences between our result and the WIBA predictions.  
	 
\subsection{Quantum annealing}\label{subsec:quantum-annealing}
	 	
	    As a final step, we focus here on a typical quantum annealing problem, whose time-dependent Hamiltonian is built from Eq.~\eqref{eq:tls-hamiltonian} using a linear interpolating schedule, \ie, $ \Gamma(t) = (1 - t/\tf) \Gamma $ and $ \epsilon(t) = \epsilon t / \tf $. For any fixed final annealing time, the time-dependent Hamiltonian then reads
	 	\begin{equation}\label{eq:hamiltonian-quantum-annealing}
	 		\ham_S(\theta) = - \qty(1-\theta) \Gamma \sigma_x - \theta \epsilon \sigma_z,
	 	\end{equation}
	 	where $ \theta = t/\tf \in \rng{0}{1} $ is a dimensionless time. We choose the transverse field $ \Gamma $ as our reference energy scale and fix $ \epsilon = \Gamma $. As prescribed by AQC, we start by preparing the reduced system at $ \theta  = 0 $ in the instantaneous eigenstate of $ \ham_S(0) $, \ie, a fully displaced state having maximum kinetic energy, and we let it evolve towards the localized ground state of $ \ham_S(1) $. The environment is initialized in its thermal equilibrium state, and, in order to keep the discussion simple, we restrict to the case $ T = 0 $.
	 	
	 	With our SIL method, we are able to simulate both short and long time dynamics, while the usually employed tools for simulating AQC algorithms strictly require long annealing times ($ \tf\to\infty $) in order to provide reliable results. Computational efforts scale linearly with the final annealing time in both cases. Among these tools, the Lindblad equation~\eqref{eq:lindblad-equation} for the reduced ground state occupation probability can be solved analytically in the adiabatic limit~\cite{albash:decoherence}, and provides the solution, in the instantaneous eigenbasis of $ \ham_S(\theta) $,
	 	\begin{equation}\label{eq:quantum-annealing-lindblad-fidelity}
	 		\rho_{--}(\theta) = \frac{1}{G(\theta)} \qty[\rho_{--}(0) + \int_{0}^{\theta} F(\theta') G(\theta') \dd{\theta'}],
	 	\end{equation}
	 	where, at zero temperature,
	 	\begin{gather}
	 		G(\theta) = \exp\int_{0}^{\theta} F(\theta') \dd{\theta'},\\
	 		F(\theta) = \tf \, \xi^2(\theta) \gamma\qty(\Delta(\theta)),\\
	 		\xi(\theta) = 2\Gamma\frac{1-\theta}{\Delta(\theta)},
	 	\end{gather}
	 	and $ \Delta(\theta) $ is the instantaneous reduced spectral gap. 
	 	Eq.~\eqref{eq:quantum-annealing-lindblad-fidelity} predicts that, at long $ \tf $, the fidelity saturates to $ \rho_{--} = 1 $ independently of the system-bath coupling strength $ \eta $, which only affects the characteristic relaxation time, proportional to $ \eta^{-1} $. This reflects the Born-Markov approximation: the bath state is uncorrelated from the reduced system state, hence, in this picture, the only effect of the zero-temperature reservoir is to drive the TLS towards its ground state. However, as discussed in Sec.~\ref{subsec:spin-boson}, in IC this picture is misleading as entangled system-bath states may arise, significantly modifying the occupations of the qubit eigenstates.
	 	
\begin{figure}[tb]
	 		\centering
	 		\includegraphics[width=\linewidth]{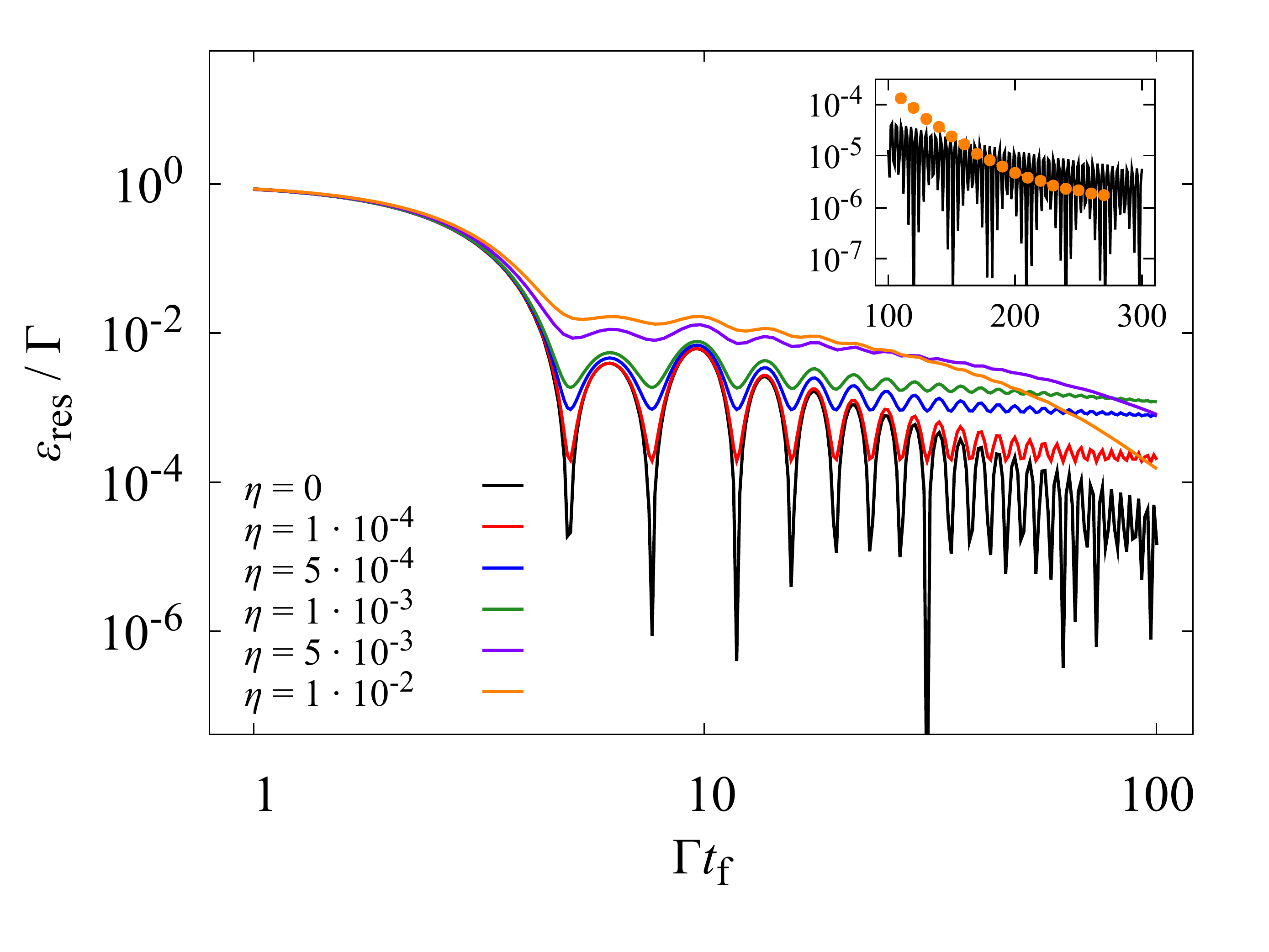}
	 		\caption{Residual energy, in units $ \Gamma $, as a function of the final annealing time, in units $ 1/\Gamma $, for coupling strengths $ \eta $ going from zero to $ 10^{-2} $. Simulations involve $ M = 200 $ bosonic modes at $ T = 0 $, with $ \omega\ped{c} = 10\Gamma $ and $ N\ped{ph} = 3 $ (Ohmic bath). At short times, the environment does not have the time to act and the system always stays close to the isolated solution. At intermediate times, the residual energy shows a plateau. At longer times, a further decrease of $ \epsilon\ped{res} $ brings the solution again towards the isolated case. At these time scales, this effect is visible only at IC. The inset focuses on longer annealing times (reached using $ M = 450 $ modes) and follows the same color scheme as the main plot. Here, we show that at long times the effect of the bath may be beneficial for the annealing.}
	 		\label{fig:quantum-annealing}
	 	\end{figure}

\begin{figure}[tb]
	 		\centering
	 		\includegraphics[width=\linewidth]{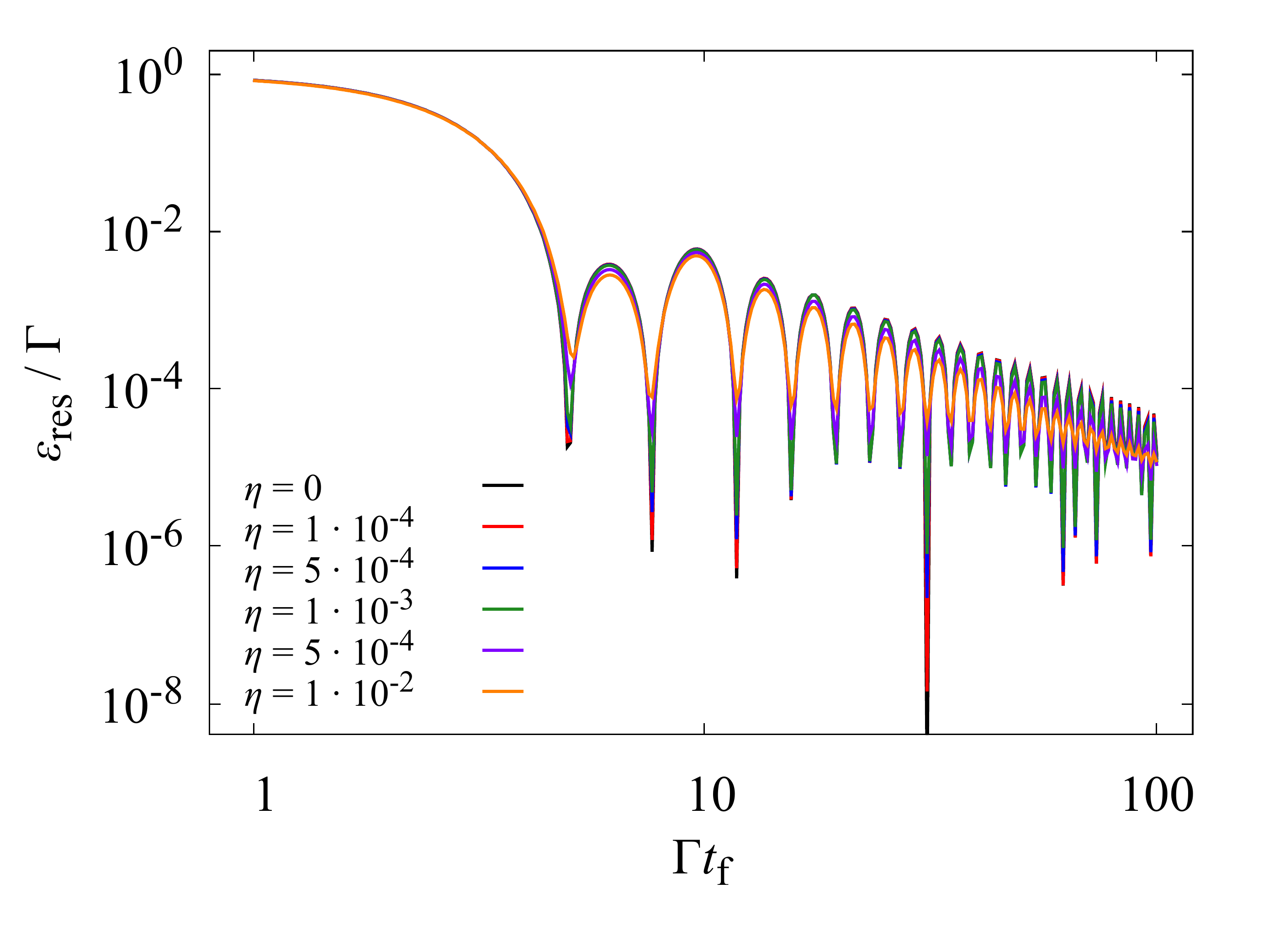}
	 		\caption{Residual energy, in units $ \Gamma $, as a function of the final annealing time, in units $ 1/\Gamma $, for coupling strengths $ \eta $ going from zero to $ 10^{-2} $, simulated using the Lindblad master equation at $ T = 0 $ and a cut-off frequency $ \omega\ped{c} = 10\Gamma $ (Ohmic bath). The only noticeable effect of increasing $ \eta $ is the progressive damping of short time oscillations in the ground state occupation, but the plateau and the following decrease of $ \epsilon\ped{res} $ observed in Fig.~\ref{fig:quantum-annealing} are not recovered.}
	 		\label{fig:quantum-annealing-lindblad}
	 	\end{figure}

	 	In order to catch the correct physics in this interesting regime, we coupled this system to $ M = 200 $ modes, each possibly occupied by maximum $ N\ped{ph} = 3 $ phonons. At $ \theta = 1 $, we measured the excess energy $ \epsilon\ped{res} $ with respect to the reduced ground state energy $ \epsilon\ped{gs} = -\epsilon $. For a TLS, $ \epsilon\ped{res} $ is proportional to the ground state error $ 1 - \rho_{--} $, \ie,
	 	\begin{equation}\label{eq:residual-energy}
	 		\epsilon\ped{res} \equiv \tr[\ham_S(1) \rho_S(1)] - \epsilon\ped{gs} = 2 \epsilon \qty[1 - \rho_{--}(1)].
	 	\end{equation}
	 	In Fig.~\ref{fig:quantum-annealing}, we compare the residual energy, in units $ \Gamma $, as a function of the final annealing time (in units $ 1/\Gamma $) of several TLSs, coupled with different coupling constants to an Ohmic environment. Similar curves, obtained by numerical integration of the Lindblad equation using a fourth-order Runge-Kutta routine, are shown in Fig.~\ref{fig:quantum-annealing-lindblad}. By comparing the curves, it is evident that system-bath correlations, disregarded by the Lindblad QME, modify quantitatively and also qualitatively the behavior of the solution in the analyzed time range, hence a Born-Markov dynamics is not able to reproduce the correct behavior.
	 	
	 	In fact, in the Lindblad picture of Fig.~\ref{fig:quantum-annealing-lindblad}, the only noticeable effect of progressively increasing  system-bath coupling strength is a very small damping of the amplitude of short-time oscillations in the ground state occupation, while both the function profile at short times and the long times power-law tail are preserved in presence of a dissipative environment. By contrast, what we found with our SIL method (Fig.~\ref{fig:quantum-annealing}) is that, while the description at short times is in agreement with Lindblad results, at intermediate times system-bath correlations tend to increase the value of the residual energy with respect to the isolated case $ \eta = 0 $, and this feature is not present in the QME solution. In fact, at intermediate times we observe a transient plateau, anticipating a further decrease of the residual energy towards the isolated case. As the exhaustion time of the plateau inversely depends on $ \eta $, we observe a non-monotonic behavior of the residual energy as a function of $ \eta $ in the time window where this decrease takes place. 
	 	The oscillations in the residual energy of the closed system are well-known and due to the finite annealing time of the chosen schedule. The effect of the environment is to suppress these oscillations in the open system case. Moreover, the residual energy of the open system can become smaller than its closed system counterpart, but this effect can only occur at some particular values of the final annealing times. This is evident in the inset of Fig.~\ref{fig:quantum-annealing}, showing that eventually, at longer annealing times, the curve corresponding to $ \eta = 10^{-2} $ approximately tends to the mean value of the closed system oscillation pattern. Whether or not this feature survives also at low but finite temperature is currently under investigation.
	 	
	 	As for the analysis at the end of Sec.~\ref{subsec:spin-boson}, concerning the oscillation frequency of $ \ev{\sigma_z(t)} $, it may be tempting to explain the influence of the environment on the quantum annealing in terms of a renormalization of the spectral gap of the reduced system. According to this argument, however, we should always expect a decrease of quantum annealing performances due to the gap reduction in the presence of the bath. Moreover, we should observe a progressive worsening of the annealing performances with increasing $ \eta $ due to this effect (see Eq.~\eqref{eq:renormalized-gap-nrg}). This argument reduces the full system to an effective TLS with renormalized spectral gap, which is a completely satisfactory description of time-independent problems as the SBM in Sec.~\ref{subsec:spin-boson}, but cannot rigorously reproduce the dynamical behavior of a system with time-dependent Hamiltonian. 
	 	
	 	To show that the previous picture might be misleading, recall that the physical description of a quantum annealing process can be understood in terms of the Landau-Zener (LZ) model~\cite{landau:crossings, zener:crossings}. The LZ Hamiltonian reads
	 	\begin{equation}\label{eq:landau-zener}
		 	\ham\ped{LZ}(t) = - \frac{vt}{2} \sigma_z - \frac{\Delta}{2} \sigma_x,
	 	\end{equation}
	 	it has a minimum spectral gap $ \Delta $ at $ t = 0 $ and in this case the TLS system evolves from $ t = -\infty $ to $ t = +\infty $ with sweep velocity $ v $. The adiabatic limit holds when $ v\to0 $. It has been shown in many works~\cite{thorwart1, thorwart4, wubs:lz-zero-t} that a zero-temperature thermal bath longitudinally coupled (\ie, via $ \sigma_z $) to the LZ system cannot provide any thermal speed-up with respect to the isolated dynamics for any sweep velocity $ v $, \ie, the probability of finding the system in its ground state at $ t = +\infty $ coincides with that of the closed system and is $ \eta $-independent. This exact result holds exclusively for an evolution from $ t = -\infty $ to $ t = +\infty $ and proves that, even though a renormalization of the minimal gap occurs, this does not necessarily lead to a decrease in the annealing performances. 
	 	
	 	Our Hamiltonian~\eqref{eq:hamiltonian-quantum-annealing} inherently differs from the LZ model~\eqref{eq:landau-zener}, thus the aforementioned theorem does not apply here. The LZ sweep velocity is inversely proportional to our final annealing time $ \tf $, which is always finite: $ v \propto 1 / \tf $. The residual energy in LZ is computed at $ t = +\infty $, while we always calculate it at $ t = \tf < \infty $. The finiteness of $ \tf $, experimentally more realistic than the limit $ \tf \to \infty $, is responsible for the oscillations of the residual energy in the isolated case; as a consequence, the residual energy in the open quantum annealing of the system Hamiltonian~\eqref{eq:hamiltonian-quantum-annealing} can be inside these oscillations at long times, causing a ``partial speed-up'' (\ie, occurring only at specific final annealing times $ \tf $) of the annealing procedure due to the environment. 
 	
	 \section{Conclusions}\label{sec:conclusions}
	    
	    In this work, we showed that a numerical technique based on the iterative application of the time evolution operator, obtained by an appropriate reduction of the problem in the Krylov subspace, is well-suited for describing decoherence and dissipation effects in systems where a qubit interacts with an external bath. Tuning the number of bosonic modes and the corresponding maximum occupations of single-particle basis states, this technique allows the perturbative inclusion of relevant phononic processes in the dynamics of the reduced system, going beyond the single-phonon physics, and enables to correctly describe time-correlation effects owing to the bath influence, ranging from weak to intermediate couplings. We emphasize that, within the proposed approach, both the full and the reduced density operator are not affected by any limitation as in standard perturbative methods, as non-positivity or non-preserved trace.
	    
	    The conceptual simplicity or our method and the ease of its numerical implementation allow a fine control on the limitations and possible sources of errors during the numerical simulations. Further, our technique, yielding the entire wave function of the system + bath, allows the calculation of all the observables related to either one of the two subsystems, or to both of them, for all kinds of dissipations. As evident from the number of recent publications concerning the topic~\cite{thorwart2, structured-bath-1, structured-bath-2}, there is a renewed interest in understanding the physics of structured baths and thermal reservoirs in general. Thus, our technique might be a valid tool to provide some insights on this class of phenomena. In addition, our technique allows to study general many-body and time-dependent problems without modifying the structure of our code and with no loss of precision.
	    
	    With our method, we tested the limits of a description entirely based on Born-Markov hypotheses and recovered known results, providing some insights on the reliability of known analytical approximations. 
	    We claim that this technique can be useful for studying simple open quantum systems and for simulating adiabatic quantum processors, perhaps in combination with other techniques such as NRG when the complexity grows. Moreover, it can be easily extended to the study of non-equilibrium behavior of many physical systems, \eg, qubits in presence of structured baths, externally driven qubits, small clusters of spatially-correlated qubits immersed in external environments, or many-body Ising systems restricted to symmetry subspaces. 
	    
	  \section{Acknowledgments}  
	   The authors thank G.~E.~Santoro for useful discussions and encouragement.

	\appendix
	
	\section{Lindblad equations}\label{app:lindblad}
	
			In the Lindblad approach, the  density of states $ J(\omega) $ (Eq.~\eqref{eq:spectral-function}) enters the definition of other two spectral densities, $ \gamma(\omega) $ and its Hilbert transform $ S(\omega) $~\cite{albash:decoherence}, defined as
			\begin{gather}\label{eq:spectral-functions-lindblad}
			\gamma(\omega) = 
			\begin{cases}
			\displaystyle\frac{2\uppi \, J(\omega)}{1 - \eu^{-\beta \omega}},&\text{$ \omega \ge 0 $,}\\[1ex]
			\eu^{\,\beta\omega} \gamma(-\omega),&\text{$ \omega < 0 $,}
			\end{cases}\\
			S(\omega) = \mathcal{P} \int_{-\infty}^{\infty} \frac{\gamma(\omega')}{\omega - \omega'} \frac{\dd{\omega'}}{2 \uppi};
			\end{gather}
			$ \beta $ is the inverse temperature of the environment in thermal equilibrium and $ \mathcal{P} $ denotes the Cauchy principal value. In particular, $ \gamma(\omega) $ expresses the effective decay rates of the reduced system and is the Fourier transform of the bath self-correlation function $ \mathcal{B}(t) =\ev{B(t) B(0)} $, where $ B $ is defined in Eq.~\eqref{eq:interaction-hamiltonian}. The dynamical equation for the reduced density matrix $ \rho_S = \tr_B\rho $ reads
			\begin{equation}\label{eq:lindblad-equation}
			\dv{\rho_S(t)}{t} = -\iu \comm{\ham_S(t) + \ham\ped{LS}(t)}{\rho_S(t)} + \diss[\rho_S(t)],
			\end{equation}
			where $ \ham\ped{LS} $ is the Lamb shift term and $ \diss $ is the adiabatic dissipator. They are expressed in terms of Lindblad operators, which, in the instantaneous eigenbasis $ \set{\ket{\epsilon_a(t)}} $ of $ \ham_S(t) $, have the following form:
			\begin{equation}\label{eq:lindblad-operators}
			L_{\omega}(t) = \sum_{\epsilon_b(t) - \epsilon_a(t) = \omega}\ket{\epsilon_a(t)} \mel{\epsilon_a(t)}{\sigma_z}{\epsilon_b(t)} \bra{\epsilon_b(t)}.
			\end{equation}
			In terms of Lindblad operators, $ \ham\ped{LS} $ and $ \diss $ are expressed as follows:
			\begin{gather}\label{eq:Dissipator}
			\ham\ped{LS} = \sum_{\omega} S(\omega) L^\dagger_\omega L_\omega,\\
			\diss[\rho_S] = \sum_{\omega} \gamma(\omega) \qty( L_\omega \rho_S L^\dagger_\omega - \frac{1}{2} \acomm{L^\dagger_\omega L_\omega}{\rho_S});
			\end{gather}
			we have omitted the time-dependence from frequencies and operators for shortness.
		
			\section{On the pure decoherence model}\label{app:decoherence}
			
			At $ \beta\to\infty $, the decoherence function $ K(t, \infty) $ of Eq.~\eqref{eq:exact-decoherence} is analytical for the three considered values of $ s $:
			\begin{equation}\label{eq:exact-decoherence-function-beta-infty}
				\begin{gathered}
					K(t,\infty)_{s=1/2} = 8 \qty[-1 + \cos(\omega\ped{c} t) + \sqrt{2\uppi \omega\ped{c} t} \Sf\qty(\sqrt{\frac{ 2\omega\ped{c} t }{\uppi}})];\\
					K(t,\infty)_{s=1} = 4 \qty[\upgamma - \Ci(\omega\ped{c} t) + \log(\omega\ped{c} t)];\\
					K(t,\infty)_{s=2} = 4 - \frac{4 \sin(\omega\ped{c} t)}{\omega\ped{c} t};
				\end{gathered}
			\end{equation}
			$ \upgamma $ is the Euler-Mascheroni constant, $ \Sf(x) $ is the Fresnel integral and $ \Ci(x) $ is the cosine integral. On the other hand, the Lindblad solution~\eqref{eq:exact-sigmax-lindblad} yields the approximated values at zero temperature $ \set{ \gamma(0)_{s=1/2} = 0; \gamma(0)_{s=1} = 0; \gamma(0)_{s=2} = 0 } $.

\begin{figure}[tb]
					\centering
					\includegraphics[width=\linewidth]{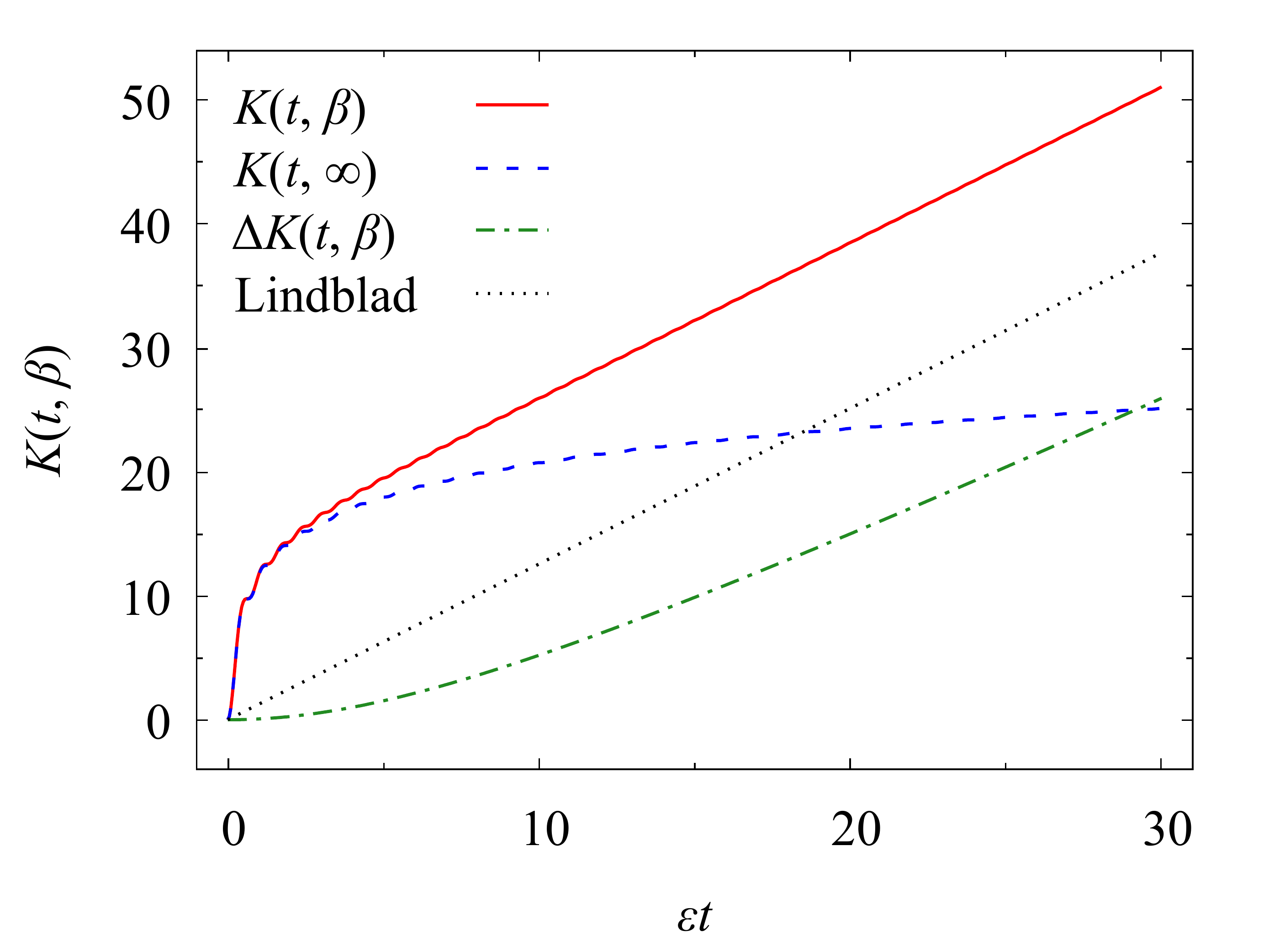}
					\caption{Decoherence function $ K(t,\beta) $ (red solid line) as a function of the dimensionless time, decomposed in its zero-temperature (blue dashed line) and finite-temperature  contributions (green dot-dashed line), for an Ohmic bath ($ s = 1 $). The black dotted line is the decoherence function predicted by the Lindblad equation. The other parameters are $ \omega\ped{c} = 10\epsilon $ and $ \beta = 10 / \epsilon $, so that $ \epsilon\tau_B = 10/\uppi $.}
					\label{fig:exact-decoherence-ohmic-omegac-10-beta-10}
			\end{figure}

\begin{figure}[tb]
				\centering
				\includegraphics[width=\linewidth]{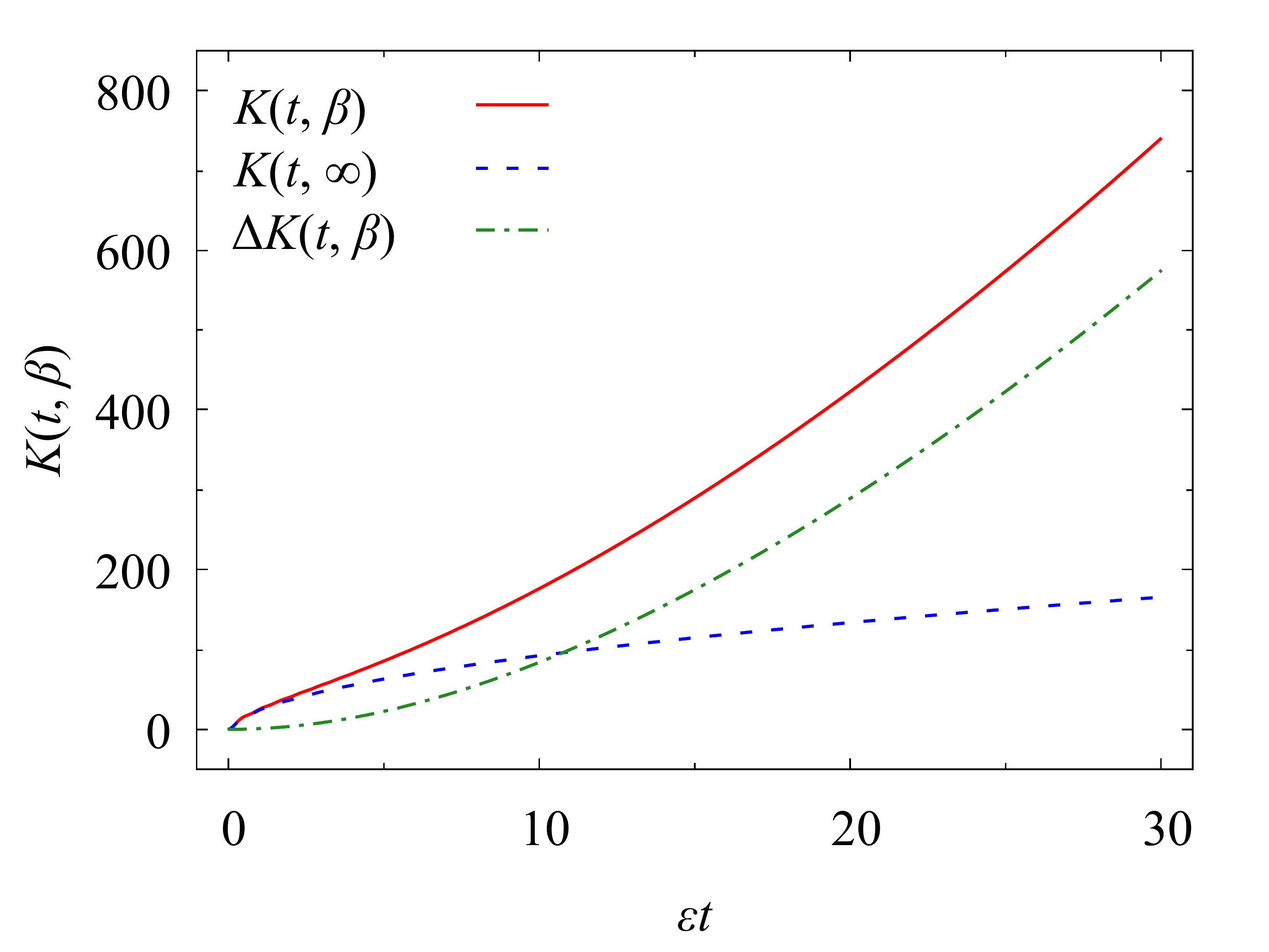}
				\caption{Same as Fig.~\ref{fig:exact-decoherence-ohmic-omegac-10-beta-10}, but for a sub-Ohmic bath ($ s = 1/2 $). The Lindblad equation predicts an infinite decoherence function (not shown).}
				\label{fig:exact-decoherence-subohmic-omegac-10-beta-10}
			\end{figure}

\begin{figure}[tb]
				\centering
				\includegraphics[width=\linewidth]{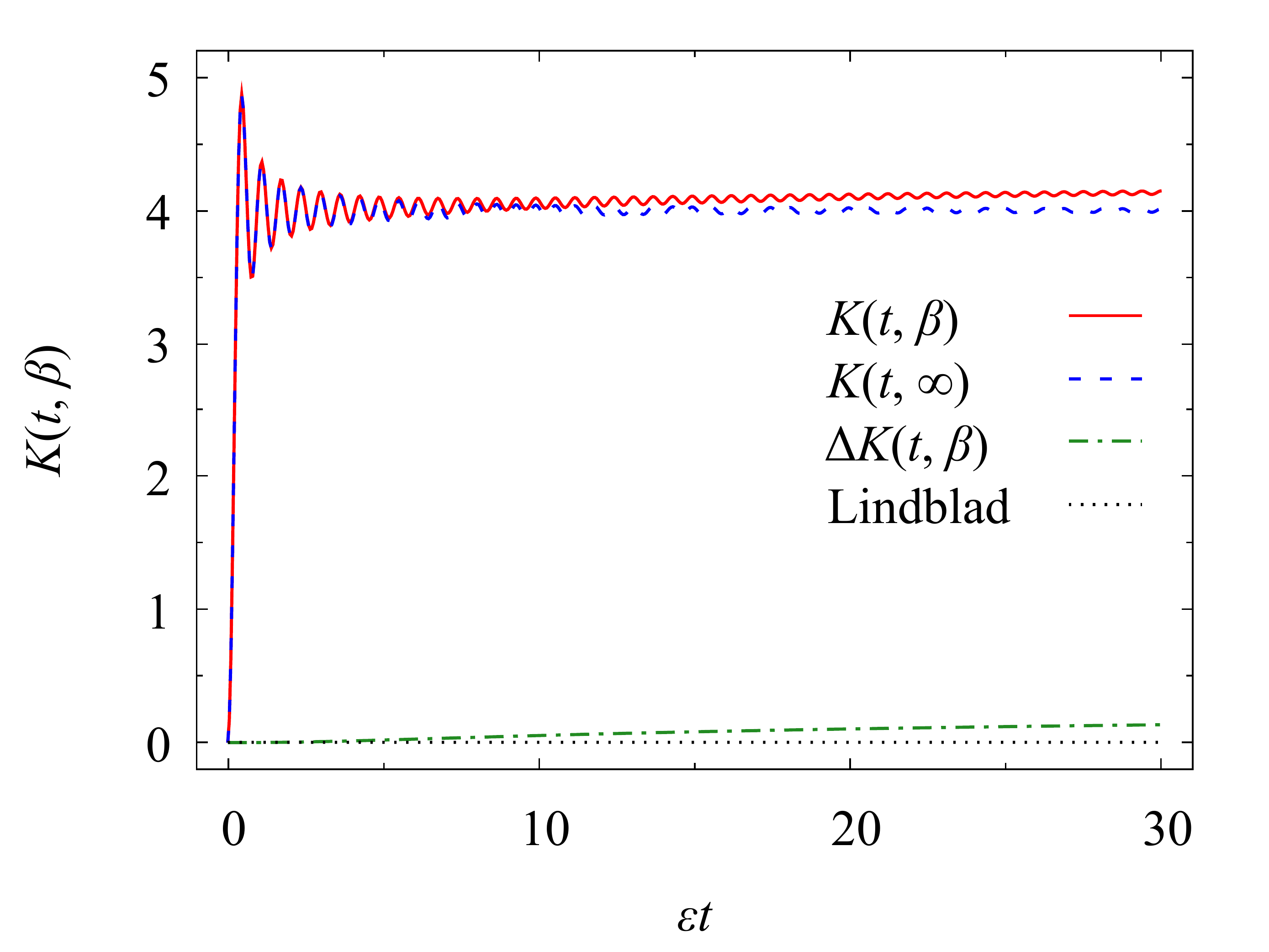}
				\caption{Same as Fig.~\ref{fig:exact-decoherence-ohmic-omegac-10-beta-10}, but for a super-Ohmic bath ($ s = 2 $). These curves show that the Lindblad theory does not include zero-temperature contributions to the decoherence function.}
				\label{fig:exact-decoherence-superohmic-omegac-10-beta-10}
			\end{figure}

			The finite temperature contribution $ \upDelta K(t,\beta) $ can be evaluated only numerically if the step-function cut-off is used in Eq.~\eqref{eq:spectral-function-continuous}. Fig.~\ref{fig:exact-decoherence-ohmic-omegac-10-beta-10} shows the behavior of $ K(t,\beta) $ and its zero- and finite-temperature contributions as a function of time (in units $ 1 / \epsilon $), for $ s = 1 $, $ \omega\ped{c} = 10 \epsilon $ and $ \beta = 10 / \epsilon $. 
			The linear trend predicted by the Lindblad equation agrees qualitatively with the finite-temperature behavior at long times of the decoherence function, but always presents a finite offset. Increasing the temperature, a regime is reached where the Lindblad equation is in good agreement with the real solution. 
			However, a further increase of the temperature leads again to discrepancies between the two models, an indication that the limit $ \beta = 0 $ 
			cannot be well-simulated by a Lindblad dynamics.  Figs.~\ref{fig:exact-decoherence-subohmic-omegac-10-beta-10} and~\ref{fig:exact-decoherence-superohmic-omegac-10-beta-10} show the same curves for $ s = 1/2 $ and $ s = 2 $, respectively.
			
		\section{SIL errors in sub- and super-Ohmic environments}\label{app:errors-sub-super}
			
			In Sec.~\ref{subsec:qubit-exact}, we showed and discussed the relative error (Eq.~\eqref{eq:relative-error}) of the SIL method for specific parameters of the simulation ($ M = 200 $, $ \omega\ped{c} = 10\epsilon $, $ T = 0 $) and in the case of an Ohmic dissipation, with respect to the analytical solution~\eqref{eq:exact-sigmax-mean}. Here, we want to perform an analogous analysis using the same parameters to simulate sub-Ohmic and super-Ohmic environments in interaction with our qubit.
			
\begin{figure}[tb]
				\centering
				\includegraphics[width=\linewidth]{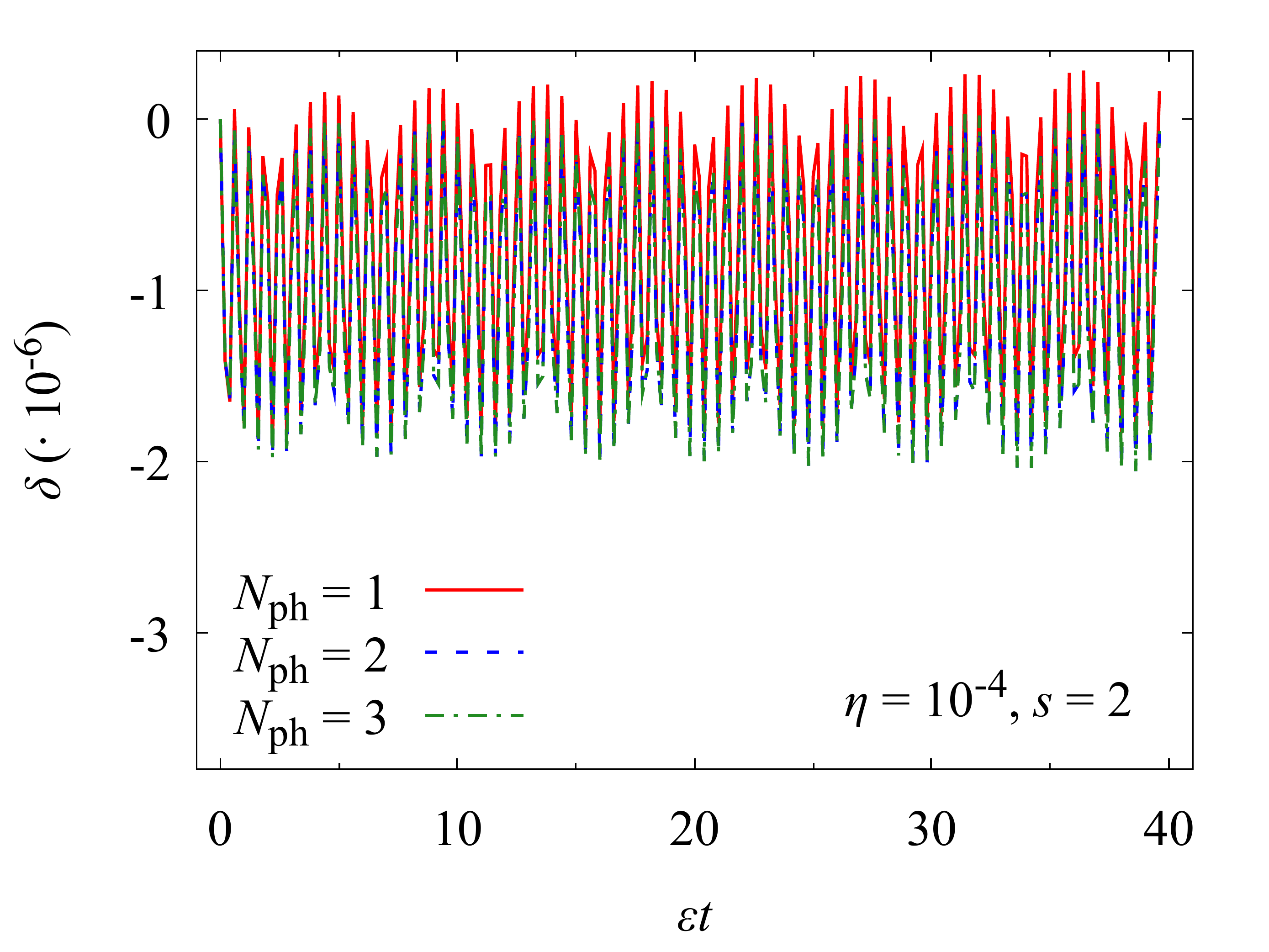}
				\caption{Relative error in the SIL approximation of $ \ev{\sigma_x(t)} $ with respect to the analytical solution, for a super-Ohmic bath ($ s = 2 $) at $ T = 0 $, coupled with $ \eta = 10^{-4} $. }
				\label{fig:exact200modierroresuperohmiceta00001nph123}
			\end{figure}

\begin{figure}[tb]
				\centering
				\includegraphics[width=\linewidth]{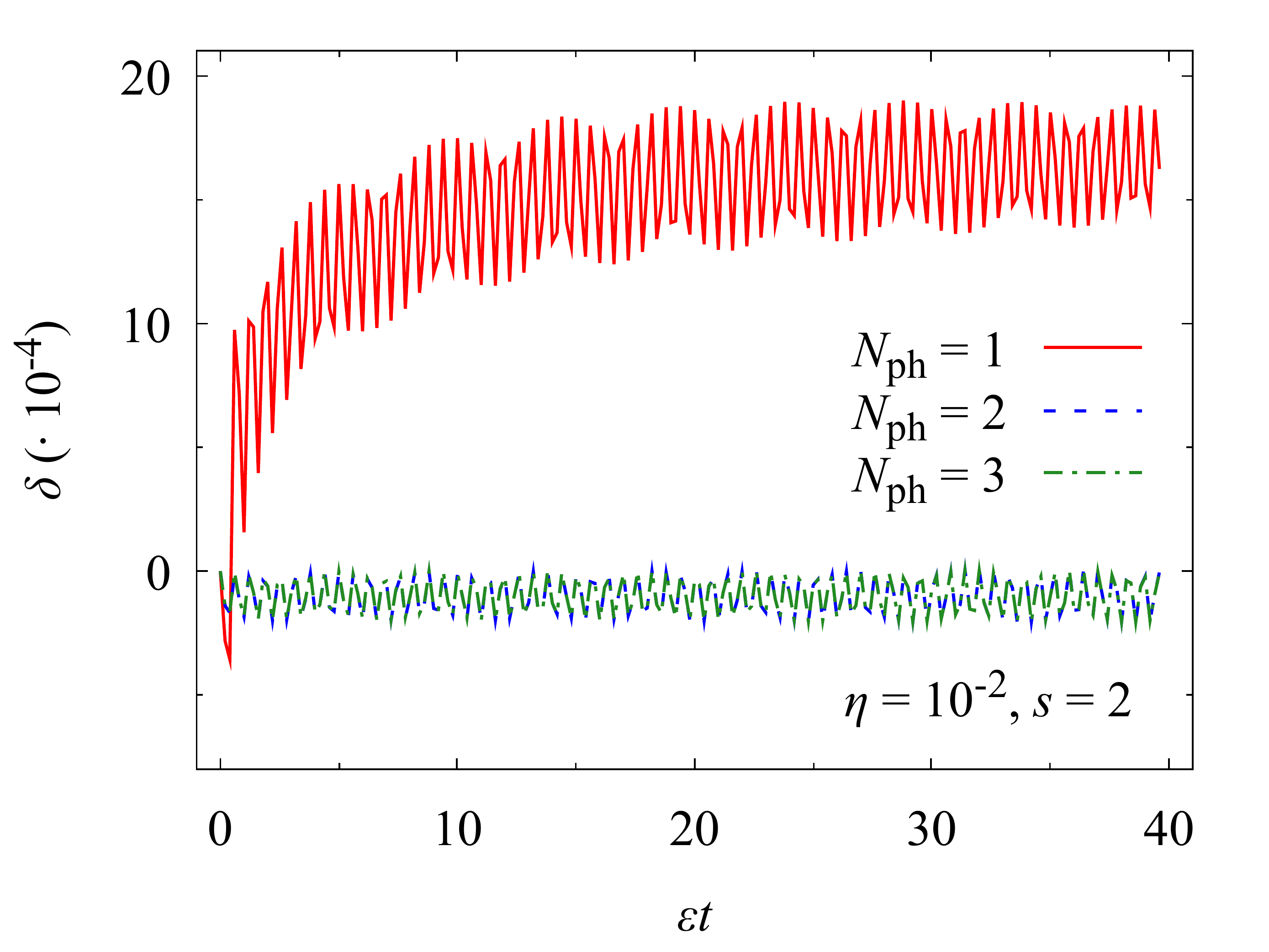}
				\caption{Relative error in the SIL approximation of $ \ev{\sigma_x(t)} $ with respect to the analytical solution, for an super-Ohmic bath ($ s = 2 $) at $ T = 0 $, coupled with $ \eta = 10^{-2} $. }
				\label{fig:exact200modierroresuperohmiceta001nph123}
			\end{figure}

			Figs.~\ref{fig:exact200modierroresuperohmiceta00001nph123} and~\ref{fig:exact200modierroresuperohmiceta001nph123} show the errors in WC and IC, respectively, for the super-Ohmic bath with $ s = 2 $. The relative error in this case is about one order of magnitude lower than the Ohmic case of Fig.~\ref{fig:exact200modierroreohmiceta00001nph123}, at equal parameters, \ie, convergence is faster for super-Ohmic environments.
			
\begin{figure}[tb]
				\centering
				\includegraphics[width=\linewidth]{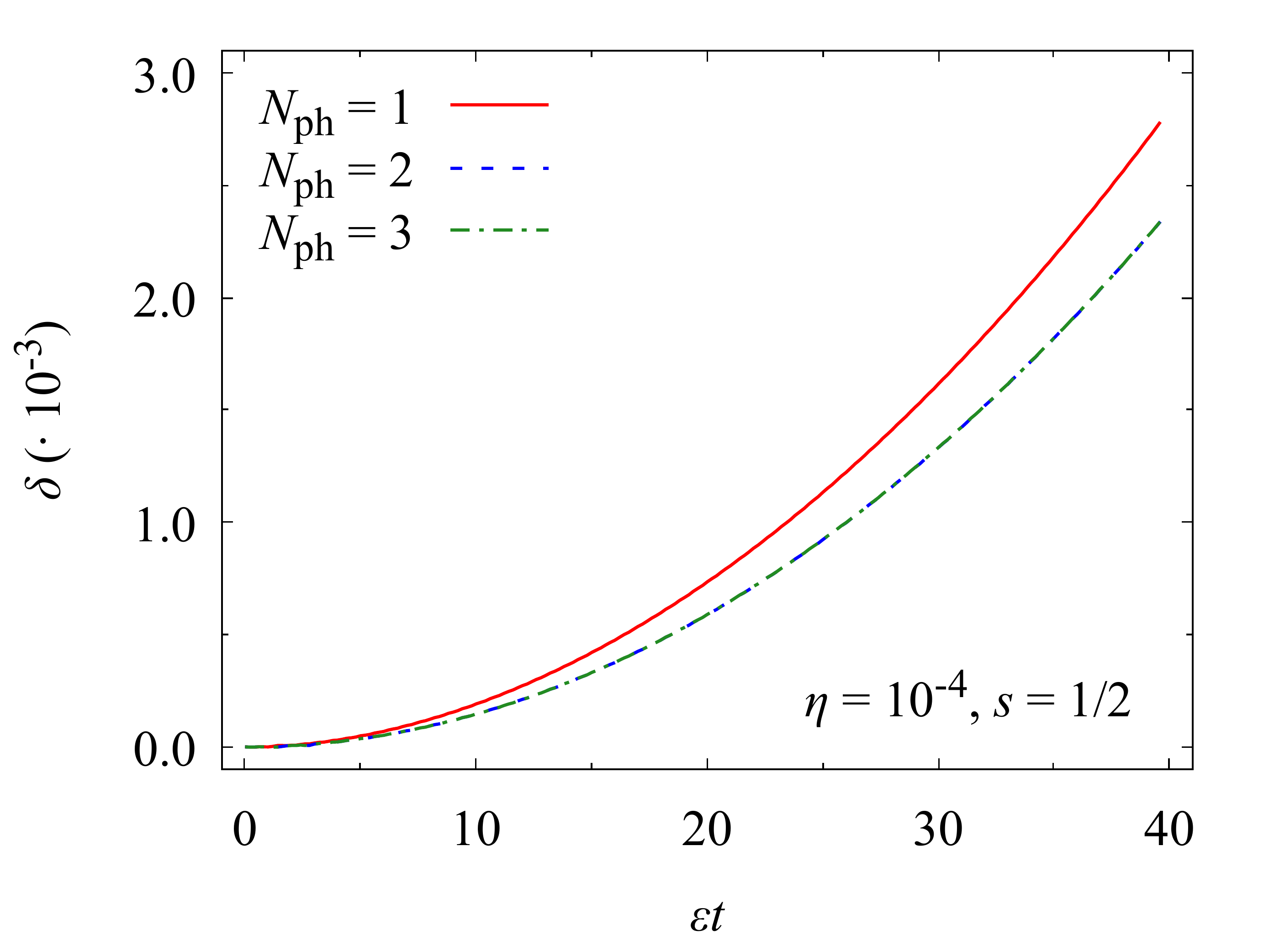}
				\caption{Relative error in the SIL approximation of $ \ev{\sigma_x(t)} $ with respect to the analytical solution, for a sub-Ohmic bath ($ s = 1/2 $) at $ T = 0 $, coupled with $ \eta = 10^{-4} $. }
				\label{fig:exact200modierroresubohmiceta00001nph123}
			\end{figure}

\begin{figure}[tb]
				\centering
				\includegraphics[width=\linewidth]{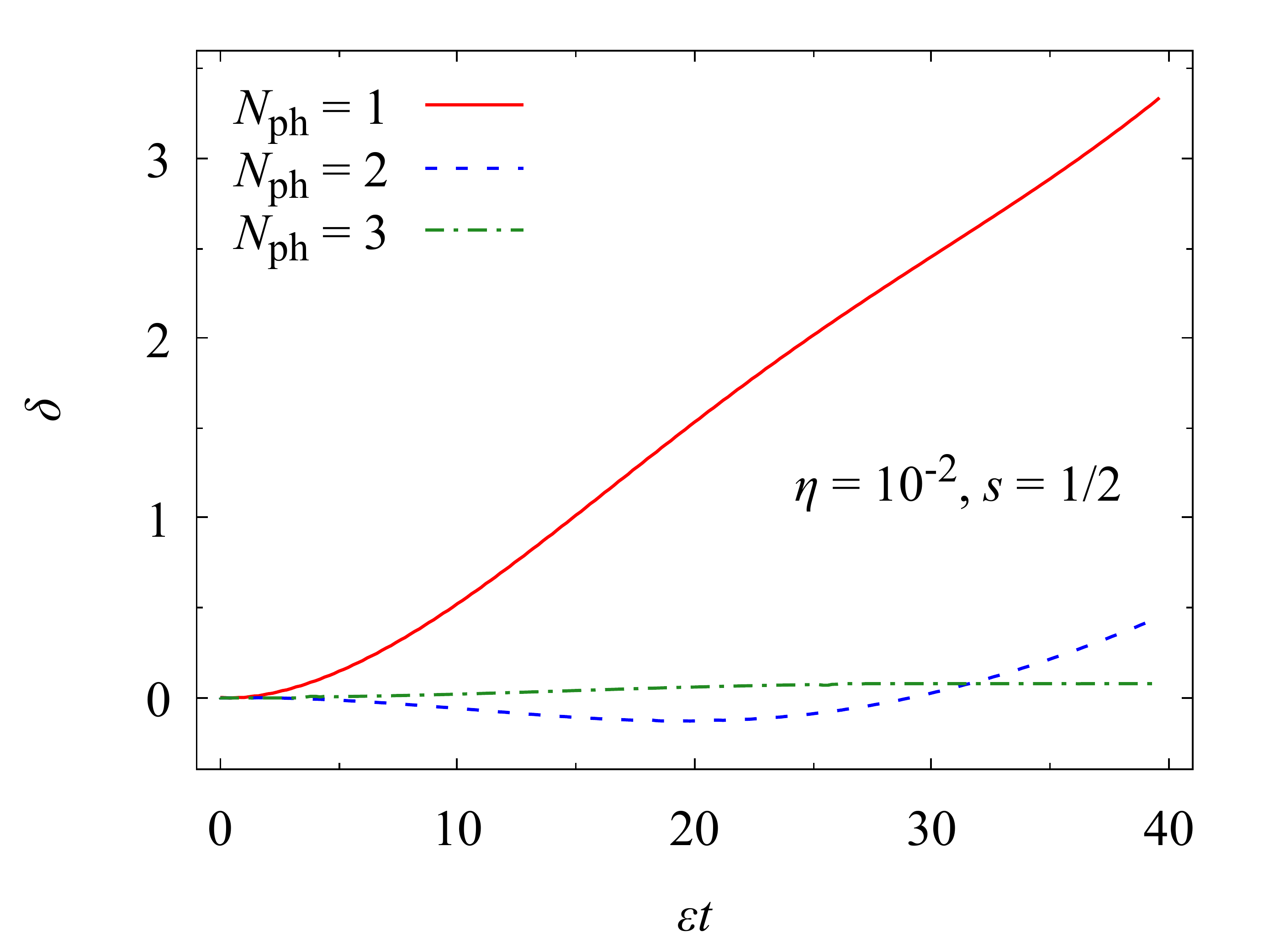}
				\caption{Relative error in the SIL approximation of $ \ev{\sigma_x(t)} $ with respect to the analytical solution, for a sub-Ohmic bath ($ s = 1/2 $) at $ T = 0 $, coupled with $ \eta = 10^{-2} $. }
				\label{fig:exact200modierroresubohmiceta001nph123}
			\end{figure}

			On the contrary, sub-Ohmic baths show a slower convergence rate to the real solution, and this is most likely due to the functional form of their spectral function. A uniform sampling does not take into account the abundance of low-frequency modes with respect to high-energy ones, and this reflects on higher relative errors in the approximation if compared with (super-)Ohmic baths (see Figs.~\ref{fig:exact200modierroresubohmiceta00001nph123} and~\ref{fig:exact200modierroresubohmiceta001nph123}). We stress that this is not a limitation of our method, as it can be easily circumvented by recurring to alternative sampling schedules, focusing on the low-frequency part of the bosonic spectrum. We also underline that the WC regime is well-reproduced even with a uniform sampling and only one phononic excitation, while, at intermediate couplings, multiple-phonon processes are strictly needed.
			
		\section{Spin-boson model in the Lindblad approximation}\label{app:spinboslind}
		
		An analytical expression for the reduced density matrix of the SBM in the Born-Markov approximation can be derived by using Eq.~\eqref{eq:lindblad-equation}. We restrict to the unbiased case, thus the qubit Hamiltonian reads $ H_S = -\Gamma \sigma_x $. By choosing the eigenstates $ \ket{\hat{x}; \pm} $ of $\ham_S$ as a set of basis states, and fixing the values of the reduced density matrix $\qty[\rho_{S} (0)]_{ij}$, $i,j=\pm $ at initial time $t=0$, the Lindblad solution for the considered expectation values reads~\cite{albash:decoherence}                   
		\begin{equation}\label{eq:densitylindsbm}
		\begin{gathered}
		\rho_{-+}(t) = \rho_{-+}(0) \eu^{-\iu\qty[S\qty(2\Gamma) - S\qty(-2\Gamma) + 2\Gamma] \, t} \eu^{-t/T_{2}};\\
		\rho_{--}(t) = \rho\ped{G}(-) + \qty[\rho_{--}(0) - \rho\ped{G}(-)] \eu^{-t/T_{1}},
		\end{gathered}
		\end{equation}
		where $\rho\ped{G}(\pm) = \eu^{\pm \beta\Gamma} / \partitionFunction $ are the Gibbs distributions associated with the eigenstates $ \ket{\hat{x}; \pm} $, respectively, $ \partitionFunction = \eu^{ \beta\Gamma} + \eu^{- \beta\Gamma} $ is the partition function, $S(\omega)$ is defined in Eq.~\eqref{eq:spectral-functions-lindblad} and the times $ T_{1,  2} $ are equal to
		\begin{align}\label{eq:decohtimes}
		T_{1} = \frac{1}{\gamma(2\Gamma)\qty(1+\eu^{-2\beta \Gamma})}, \qquad T_{2} = 2T_{1}.
		\end{align}
		Starting from Eq.~\eqref{eq:densitylindsbm}, the time evolution for the expectation values of $\ev{\sigma_x(t)}$ and $\ev{\sigma_z(t)}$ can be derived:
		\begin{equation}\label{eq:Lindspinboson}
		\begin{gathered}
		\ev{\sigma_z(t)}\ped{L} = \rho_{-+}(0) \eu^{-\iu\qty[S\qty(2\Gamma) - S\qty(-2\Gamma) + 2\Gamma] \, t} \eu^{-t/T_{2}}  + \, \cc;\\
		\ev{\sigma_x(t)}\ped{L} =  \tanh(\beta\Gamma) -2\qty[\rho_{--}(0) - \rho\ped{G}(-)] \eu^{-t/T_{1}}.
		\end{gathered}
		\end{equation}
		As evident from Eq.~\eqref{eq:Lindspinboson}, the asymptotic value of $\ev{\sigma_x(t)}\ped{L}$ does not depend on the coupling strength $\eta$, but the latter affects only the equilibration time $ T_{1} $.


\begin{thebibliography}{51}%
		\makeatletter
		\providecommand \@ifxundefined [1]{%
			\@ifx{#1\undefined}
		}%
		\providecommand \@ifnum [1]{%
			\ifnum #1\expandafter \@firstoftwo
			\else \expandafter \@secondoftwo
			\fi
		}%
		\providecommand \@ifx [1]{%
			\ifx #1\expandafter \@firstoftwo
			\else \expandafter \@secondoftwo
			\fi
		}%
		\providecommand \natexlab [1]{#1}%
		\providecommand \enquote  [1]{``#1''}%
		\providecommand \bibnamefont  [1]{#1}%
		\providecommand \bibfnamefont [1]{#1}%
		\providecommand \citenamefont [1]{#1}%
		\providecommand \href@noop [0]{\@secondoftwo}%
		\providecommand \href [0]{\begingroup \@sanitize@url \@href}%
		\providecommand \@href[1]{\@@startlink{#1}\@@href}%
		\providecommand \@@href[1]{\endgroup#1\@@endlink}%
		\providecommand \@sanitize@url [0]{\catcode `\\12\catcode `\$12\catcode
			`\&12\catcode `\#12\catcode `\^12\catcode `\_12\catcode `\%12\relax}%
		\providecommand \@@startlink[1]{}%
		\providecommand \@@endlink[0]{}%
		\providecommand \url  [0]{\begingroup\@sanitize@url \@url }%
		\providecommand \@url [1]{\endgroup\@href {#1}{\urlprefix }}%
		\providecommand \urlprefix  [0]{URL }%
		\providecommand \Eprint [0]{\href }%
		\providecommand \doibase [0]{http://dx.doi.org/}%
		\providecommand \selectlanguage [0]{\@gobble}%
		\providecommand \bibinfo  [0]{\@secondoftwo}%
		\providecommand \bibfield  [0]{\@secondoftwo}%
		\providecommand \translation [1]{[#1]}%
		\providecommand \BibitemOpen [0]{}%
		\providecommand \bibitemStop [0]{}%
		\providecommand \bibitemNoStop [0]{.\EOS\space}%
		\providecommand \EOS [0]{\spacefactor3000\relax}%
		\providecommand \BibitemShut  [1]{\csname bibitem#1\endcsname}%
		\let\auto@bib@innerbib\@empty
		\bibitem [{\citenamefont {Caldeira}\ and\ \citenamefont
			{Leggett}(1983)}]{caldeira:caldeira-leggett}%
		\BibitemOpen
		\bibfield  {author} {\bibinfo {author} {\bibfnamefont {A.~O.}\ \bibnamefont
				{Caldeira}}\ and\ \bibinfo {author} {\bibfnamefont {A.~J.}\ \bibnamefont
				{Leggett}},\ }\href
		{http://www.sciencedirect.com/science/article/pii/0003491683902026}
		{\bibfield  {journal} {\bibinfo  {journal} {Annals of Physics}\ }\textbf
			{\bibinfo {volume} {149}},\ \bibinfo {pages} {374} (\bibinfo {year}
			{1983})}\BibitemShut {NoStop}%
		\bibitem [{\citenamefont {Breuer}\ and\ \citenamefont
			{Petruccione}(2007)}]{breuer:open-quantum}%
		\BibitemOpen
		\bibfield  {author} {\bibinfo {author} {\bibfnamefont {H.~P.}\ \bibnamefont
				{Breuer}}\ and\ \bibinfo {author} {\bibfnamefont {F.}~\bibnamefont
				{Petruccione}},\ }\href@noop {} {\emph {\bibinfo {title} {The Theory of Open
					Quantum Systems}}}\ (\bibinfo  {publisher} {OUP Oxford},\ \bibinfo {year}
		{2007})\BibitemShut {NoStop}%
		\bibitem [{\citenamefont {Egger}\ and\ \citenamefont {Mak}(1994)}]{PathInt1}%
		\BibitemOpen
		\bibfield  {author} {\bibinfo {author} {\bibfnamefont {R.}~\bibnamefont
				{Egger}}\ and\ \bibinfo {author} {\bibfnamefont {C.~H.}\ \bibnamefont
				{Mak}},\ }\href {\doibase 10.1103/PhysRevB.50.15210} {\bibfield  {journal}
			{\bibinfo  {journal} {Phys. Rev. B}\ }\textbf {\bibinfo {volume} {50}},\
			\bibinfo {pages} {15210} (\bibinfo {year} {1994})}\BibitemShut {NoStop}%
		\bibitem [{\citenamefont {Strathearn}\ \emph {et~al.}(2017)\citenamefont
			{Strathearn}, \citenamefont {Lovett},\ and\ \citenamefont
			{Kirton}}]{PathInt2}%
		\BibitemOpen
		\bibfield  {author} {\bibinfo {author} {\bibfnamefont {A.}~\bibnamefont
				{Strathearn}}, \bibinfo {author} {\bibfnamefont {B.~W.}\ \bibnamefont
				{Lovett}}, \ and\ \bibinfo {author} {\bibfnamefont {P.}~\bibnamefont
				{Kirton}},\ }\href {http://stacks.iop.org/1367-2630/19/i=9/a=093009}
		{\bibfield  {journal} {\bibinfo  {journal} {New Journal of Physics}\ }\textbf
			{\bibinfo {volume} {19}},\ \bibinfo {pages} {093009} (\bibinfo {year}
			{2017})}\BibitemShut {NoStop}%
		\bibitem [{\citenamefont {Topaler}\ and\ \citenamefont {Makri}(1993)}]{quapi}%
		\BibitemOpen
		\bibfield  {author} {\bibinfo {author} {\bibfnamefont {M.}~\bibnamefont
				{Topaler}}\ and\ \bibinfo {author} {\bibfnamefont {N.}~\bibnamefont
				{Makri}},\ }\href {\doibase https://doi.org/10.1016/0009-2614(93)89135-5}
		{\bibfield  {journal} {\bibinfo  {journal} {Chemical Physics Letters}\
			}\textbf {\bibinfo {volume} {210}},\ \bibinfo {pages} {285 } (\bibinfo {year}
			{1993})}\BibitemShut {NoStop}%
		\bibitem [{\citenamefont {Dattani}(2013{\natexlab{a}})}]{weenie-dattani}%
		\BibitemOpen
		\bibfield  {author} {\bibinfo {author} {\bibfnamefont {N.~S.}\ \bibnamefont
				{Dattani}},\ }\href {\doibase https://doi.org/10.1016/j.cpc.2013.07.001}
		{\bibfield  {journal} {\bibinfo  {journal} {Computer Physics Communications}\
			}\textbf {\bibinfo {volume} {184}},\ \bibinfo {pages} {2828 } (\bibinfo
			{year} {2013}{\natexlab{a}})}\BibitemShut {NoStop}%
		\bibitem [{\citenamefont {Dattani}(2013{\natexlab{b}})}]{weenie-dattani-2}%
		\BibitemOpen
		\bibfield  {author} {\bibinfo {author} {\bibfnamefont {N.~S.}\ \bibnamefont
				{Dattani}},\ }\href {http://dx.doi.org/10.6084/m9.figshare.823549} {\enquote
			{\bibinfo {title} {Feyndyn},}\ } (\bibinfo {year}
		{2013}{\natexlab{b}})\BibitemShut {NoStop}%
		\bibitem [{\citenamefont {Leggett}\ \emph {et~al.}(1987)\citenamefont
			{Leggett}, \citenamefont {Chakravarty}, \citenamefont {Dorsey}, \citenamefont
			{Fisher}, \citenamefont {Garg},\ and\ \citenamefont {Zwerger}}]{Leggett}%
		\BibitemOpen
		\bibfield  {author} {\bibinfo {author} {\bibfnamefont {A.~J.}\ \bibnamefont
				{Leggett}}, \bibinfo {author} {\bibfnamefont {S.}~\bibnamefont
				{Chakravarty}}, \bibinfo {author} {\bibfnamefont {A.~T.}\ \bibnamefont
				{Dorsey}}, \bibinfo {author} {\bibfnamefont {M.~P.~A.}\ \bibnamefont
				{Fisher}}, \bibinfo {author} {\bibfnamefont {A.}~\bibnamefont {Garg}}, \ and\
			\bibinfo {author} {\bibfnamefont {W.}~\bibnamefont {Zwerger}},\ }\href
		{\doibase 10.1103/RevModPhys.59.1} {\bibfield  {journal} {\bibinfo  {journal}
				{Rev. Mod. Phys.}\ }\textbf {\bibinfo {volume} {59}},\ \bibinfo {pages} {1}
			(\bibinfo {year} {1987})}\BibitemShut {NoStop}%
		\bibitem [{\citenamefont {Nesi}\ \emph {et~al.}(2007)\citenamefont {Nesi},
			\citenamefont {Paladino}, \citenamefont {Thorwart},\ and\ \citenamefont
			{Grifoni}}]{Grifoni}%
		\BibitemOpen
		\bibfield  {author} {\bibinfo {author} {\bibfnamefont {F.}~\bibnamefont
				{Nesi}}, \bibinfo {author} {\bibfnamefont {E.}~\bibnamefont {Paladino}},
			\bibinfo {author} {\bibfnamefont {M.}~\bibnamefont {Thorwart}}, \ and\
			\bibinfo {author} {\bibfnamefont {M.}~\bibnamefont {Grifoni}},\ }\href
		{\doibase 10.1103/PhysRevB.76.155323} {\bibfield  {journal} {\bibinfo
				{journal} {Phys. Rev. B}\ }\textbf {\bibinfo {volume} {76}},\ \bibinfo
			{pages} {155323} (\bibinfo {year} {2007})}\BibitemShut {NoStop}%
		\bibitem [{\citenamefont {Hur}(2008)}]{LeHurNRG1}%
		\BibitemOpen
		\bibfield  {author} {\bibinfo {author} {\bibfnamefont {K.~L.}\ \bibnamefont
				{Hur}},\ }\href {\doibase https://doi.org/10.1016/j.aop.2007.12.003}
		{\bibfield  {journal} {\bibinfo  {journal} {Annals of Physics}\ }\textbf
			{\bibinfo {volume} {323}},\ \bibinfo {pages} {2208 } (\bibinfo {year}
			{2008})}\BibitemShut {NoStop}%
		\bibitem [{\citenamefont {Le~Hur}\ \emph {et~al.}(2007)\citenamefont {Le~Hur},
			\citenamefont {Doucet-Beaupr\'e},\ and\ \citenamefont
			{Hofstetter}}]{LeHurImpurity}%
		\BibitemOpen
		\bibfield  {author} {\bibinfo {author} {\bibfnamefont {K.}~\bibnamefont
				{Le~Hur}}, \bibinfo {author} {\bibfnamefont {P.}~\bibnamefont
				{Doucet-Beaupr\'e}}, \ and\ \bibinfo {author} {\bibfnamefont
				{W.}~\bibnamefont {Hofstetter}},\ }\href {\doibase
			10.1103/PhysRevLett.99.126801} {\bibfield  {journal} {\bibinfo  {journal}
				{Phys. Rev. Lett.}\ }\textbf {\bibinfo {volume} {99}},\ \bibinfo {pages}
			{126801} (\bibinfo {year} {2007})}\BibitemShut {NoStop}%
		\bibitem [{\citenamefont {Lindblad}(1976)}]{lindblad1976}%
		\BibitemOpen
		\bibfield  {author} {\bibinfo {author} {\bibfnamefont {G.}~\bibnamefont
				{Lindblad}},\ }\href {https://projecteuclid.org:443/euclid.cmp/1103899849}
		{\bibfield  {journal} {\bibinfo  {journal} {Comm. Math. Phys.}\ }\textbf
			{\bibinfo {volume} {48}},\ \bibinfo {pages} {119} (\bibinfo {year}
			{1976})}\BibitemShut {NoStop}%
		\bibitem [{\citenamefont {Gorini}\ \emph {et~al.}(1976)\citenamefont {Gorini},
			\citenamefont {Kossakowski},\ and\ \citenamefont
			{Sudarshan}}]{GoriniKossSud76}%
		\BibitemOpen
		\bibfield  {author} {\bibinfo {author} {\bibfnamefont {V.}~\bibnamefont
				{Gorini}}, \bibinfo {author} {\bibfnamefont {A.}~\bibnamefont {Kossakowski}},
			\ and\ \bibinfo {author} {\bibfnamefont {E.~C.~G.}\ \bibnamefont
				{Sudarshan}},\ }\href {\doibase 10.1063/1.522979} {\bibfield  {journal}
			{\bibinfo  {journal} {Journal of Mathematical Physics}\ }\textbf {\bibinfo
				{volume} {17}},\ \bibinfo {pages} {821} (\bibinfo {year} {1976})},\ \Eprint
		{http://arxiv.org/abs/https://aip.scitation.org/doi/pdf/10.1063/1.522979}
		{https://aip.scitation.org/doi/pdf/10.1063/1.522979} \BibitemShut {NoStop}%
		\bibitem [{\citenamefont {Kadowaki}\ and\ \citenamefont
			{Nishimori}(1998)}]{kadowaki:qa}%
		\BibitemOpen
		\bibfield  {author} {\bibinfo {author} {\bibfnamefont {T.}~\bibnamefont
				{Kadowaki}}\ and\ \bibinfo {author} {\bibfnamefont {H.}~\bibnamefont
				{Nishimori}},\ }\href {http://link.aps.org/doi/10.1103/PhysRevE.58.5355}
		{\bibfield  {journal} {\bibinfo  {journal} {Phys. Rev. E}\ }\textbf {\bibinfo
				{volume} {58}},\ \bibinfo {pages} {5355} (\bibinfo {year}
			{1998})}\BibitemShut {NoStop}%
		\bibitem [{\citenamefont {Farhi}\ \emph {et~al.}(2000)\citenamefont {Farhi},
			\citenamefont {Goldstone}, \citenamefont {Gutmann},\ and\ \citenamefont
			{Sipser}}]{farhi:quantum-computation}%
		\BibitemOpen
		\bibfield  {author} {\bibinfo {author} {\bibfnamefont {E.}~\bibnamefont
				{Farhi}}, \bibinfo {author} {\bibfnamefont {J.}~\bibnamefont {Goldstone}},
			\bibinfo {author} {\bibfnamefont {S.}~\bibnamefont {Gutmann}}, \ and\
			\bibinfo {author} {\bibfnamefont {M.}~\bibnamefont {Sipser}},\ }\href@noop {}
		{\bibfield  {journal} {\bibinfo  {journal} {arXiv preprint
					arXiv:quant-ph/0001106}\ } (\bibinfo {year} {2000})}\BibitemShut {NoStop}%
		\bibitem [{\citenamefont {Childs}\ \emph {et~al.}(2001)\citenamefont {Childs},
			\citenamefont {Farhi},\ and\ \citenamefont {Preskill}}]{childs:robustness}%
		\BibitemOpen
		\bibfield  {author} {\bibinfo {author} {\bibfnamefont {A.~M.}\ \bibnamefont
				{Childs}}, \bibinfo {author} {\bibfnamefont {E.}~\bibnamefont {Farhi}}, \
			and\ \bibinfo {author} {\bibfnamefont {J.}~\bibnamefont {Preskill}},\ }\href
		{http://link.aps.org/doi/10.1103/PhysRevA.65.012322} {\bibfield  {journal}
			{\bibinfo  {journal} {Phys. Rev. A}\ }\textbf {\bibinfo {volume} {65}},\
			\bibinfo {pages} {012322} (\bibinfo {year} {2001})}\BibitemShut {NoStop}%
		\bibitem [{\citenamefont {Harris}\ \emph {et~al.}(2011)\citenamefont {Harris},
			\citenamefont {Berkley}, \citenamefont {Johansson}, \citenamefont {Bunyk},
			\citenamefont {Chapple}, \citenamefont {Enderud}, \citenamefont {Hilton},
			\citenamefont {Karimi}, \citenamefont {Ladizinsky}, \citenamefont
			{Ladizinsky}, \citenamefont {Oh}, \citenamefont {Perminov}, \citenamefont
			{Rich}, \citenamefont {Thom}, \citenamefont {Tolkacheva}, \citenamefont
			{Truncik}, \citenamefont {Uchaikin}, \citenamefont {Wang}, \citenamefont
			{Wilson},\ and\ \citenamefont {Rose}}]{harris:d-wave}%
		\BibitemOpen
		\bibfield  {author} {\bibinfo {author} {\bibfnamefont {R.}~\bibnamefont
				{Harris}}, \bibinfo {author} {\bibfnamefont {A.~J.}\ \bibnamefont {Berkley}},
			\bibinfo {author} {\bibfnamefont {J.}~\bibnamefont {Johansson}}, \bibinfo
			{author} {\bibfnamefont {P.}~\bibnamefont {Bunyk}}, \bibinfo {author}
			{\bibfnamefont {E.~M.}\ \bibnamefont {Chapple}}, \bibinfo {author}
			{\bibfnamefont {C.}~\bibnamefont {Enderud}}, \bibinfo {author} {\bibfnamefont
				{J.~P.}\ \bibnamefont {Hilton}}, \bibinfo {author} {\bibfnamefont
				{K.}~\bibnamefont {Karimi}}, \bibinfo {author} {\bibfnamefont
				{E.}~\bibnamefont {Ladizinsky}}, \bibinfo {author} {\bibfnamefont
				{N.}~\bibnamefont {Ladizinsky}}, \bibinfo {author} {\bibfnamefont
				{T.}~\bibnamefont {Oh}}, \bibinfo {author} {\bibfnamefont {I.}~\bibnamefont
				{Perminov}}, \bibinfo {author} {\bibfnamefont {C.}~\bibnamefont {Rich}},
			\bibinfo {author} {\bibfnamefont {M.~C.}\ \bibnamefont {Thom}}, \bibinfo
			{author} {\bibfnamefont {E.}~\bibnamefont {Tolkacheva}}, \bibinfo {author}
			{\bibfnamefont {C.~J.~S.}\ \bibnamefont {Truncik}}, \bibinfo {author}
			{\bibfnamefont {S.}~\bibnamefont {Uchaikin}}, \bibinfo {author}
			{\bibfnamefont {J.}~\bibnamefont {Wang}}, \bibinfo {author} {\bibfnamefont
				{B.}~\bibnamefont {Wilson}}, \ and\ \bibinfo {author} {\bibfnamefont
				{G.}~\bibnamefont {Rose}},\ }\href {http://dx.doi.org/10.1038/nature10012}
		{\bibfield  {journal} {\bibinfo  {journal} {Nature}\ }\textbf {\bibinfo
				{volume} {473}},\ \bibinfo {pages} {194} (\bibinfo {year}
			{2011})}\BibitemShut {NoStop}%
		\bibitem [{\citenamefont {Passarelli}\ \emph {et~al.}(2018)\citenamefont
			{Passarelli}, \citenamefont {De~Filippis}, \citenamefont {Cataudella},\ and\
			\citenamefont {Lucignano}}]{Me}%
		\BibitemOpen
		\bibfield  {author} {\bibinfo {author} {\bibfnamefont {G.}~\bibnamefont
				{Passarelli}}, \bibinfo {author} {\bibfnamefont {G.}~\bibnamefont
				{De~Filippis}}, \bibinfo {author} {\bibfnamefont {V.}~\bibnamefont
				{Cataudella}}, \ and\ \bibinfo {author} {\bibfnamefont {P.}~\bibnamefont
				{Lucignano}},\ }\href {\doibase 10.1103/PhysRevA.97.022319} {\bibfield
			{journal} {\bibinfo  {journal} {Phys. Rev. A}\ }\textbf {\bibinfo {volume}
				{97}},\ \bibinfo {pages} {022319} (\bibinfo {year} {2018})}\BibitemShut
		{NoStop}%
		\bibitem [{\citenamefont {Smelyanskiy}\ \emph {et~al.}(2017)\citenamefont
			{Smelyanskiy}, \citenamefont {Venturelli}, \citenamefont {Perdomo-Ortiz},
			\citenamefont {Knysh},\ and\ \citenamefont
			{Dykman}}]{Smelyanskiy:decoherence}%
		\BibitemOpen
		\bibfield  {author} {\bibinfo {author} {\bibfnamefont {V.~N.}\ \bibnamefont
				{Smelyanskiy}}, \bibinfo {author} {\bibfnamefont {D.}~\bibnamefont
				{Venturelli}}, \bibinfo {author} {\bibfnamefont {A.}~\bibnamefont
				{Perdomo-Ortiz}}, \bibinfo {author} {\bibfnamefont {S.}~\bibnamefont
				{Knysh}}, \ and\ \bibinfo {author} {\bibfnamefont {M.}~\bibnamefont
				{Dykman}},\ }\href@noop {} {\bibfield  {journal} {\bibinfo  {journal} {Phys.
					Rev. Lett.}\ }\textbf {\bibinfo {volume} {118}},\ \bibinfo {pages} {066802}
			(\bibinfo {year} {2017})}\BibitemShut {NoStop}%
		\bibitem [{\citenamefont {Kechedzhi}\ and\ \citenamefont
			{Smelyanskiy}(2016)}]{Smelyanskiy:decoherence2}%
		\BibitemOpen
		\bibfield  {author} {\bibinfo {author} {\bibfnamefont {K.}~\bibnamefont
				{Kechedzhi}}\ and\ \bibinfo {author} {\bibfnamefont {V.~N.}\ \bibnamefont
				{Smelyanskiy}},\ }\href@noop {} {\bibfield  {journal} {\bibinfo  {journal}
				{Phys. Rev. X}\ }\textbf {\bibinfo {volume} {6}},\ \bibinfo {pages} {021028}
			(\bibinfo {year} {2016})}\BibitemShut {NoStop}%
		\bibitem [{\citenamefont {Amin}\ \emph {et~al.}(2008)\citenamefont {Amin},
			\citenamefont {Love},\ and\ \citenamefont {Truncik}}]{amin:thermal-qa}%
		\BibitemOpen
		\bibfield  {author} {\bibinfo {author} {\bibfnamefont {M.~H.~S.}\
				\bibnamefont {Amin}}, \bibinfo {author} {\bibfnamefont {P.~J.}\ \bibnamefont
				{Love}}, \ and\ \bibinfo {author} {\bibfnamefont {C.~J.~S.}\ \bibnamefont
				{Truncik}},\ }\href {\doibase 10.1103/PhysRevLett.100.060503} {\bibfield
			{journal} {\bibinfo  {journal} {Phys. Rev. Lett.}\ }\textbf {\bibinfo
				{volume} {100}},\ \bibinfo {pages} {060503} (\bibinfo {year}
			{2008})}\BibitemShut {NoStop}%
		\bibitem [{\citenamefont {Dickson}\ \emph {et~al.}(2013)\citenamefont
			{Dickson}, \citenamefont {Johnson}, \citenamefont {Amin}, \citenamefont
			{Harris}, \citenamefont {Altomare}, \citenamefont {Berkley}, \citenamefont
			{Bunyk}, \citenamefont {Cai}, \citenamefont {Chapple}, \citenamefont
			{Chavez}, \citenamefont {Cioata}, \citenamefont {Cirip}, \citenamefont
			{deBuen}, \citenamefont {Drew-Brook}, \citenamefont {Enderud}, \citenamefont
			{Gildert}, \citenamefont {Hamze}, \citenamefont {Hilton}, \citenamefont
			{Hoskinson}, \citenamefont {Karimi}, \citenamefont {Ladizinsky},
			\citenamefont {Ladizinsky}, \citenamefont {Lanting}, \citenamefont {Mahon},
			\citenamefont {Neufeld}, \citenamefont {Oh}, \citenamefont {Perminov},
			\citenamefont {Petroff}, \citenamefont {Przybysz}, \citenamefont {Rich},
			\citenamefont {Spear}, \citenamefont {Tcaciuc}, \citenamefont {Thom},
			\citenamefont {Tolkacheva}, \citenamefont {Uchaikin}, \citenamefont {Wang},
			\citenamefont {Wilson}, \citenamefont {Merali},\ and\ \citenamefont
			{Rose}}]{dickson:thermal-qa}%
		\BibitemOpen
		\bibfield  {author} {\bibinfo {author} {\bibfnamefont {N.~G.}\ \bibnamefont
				{Dickson}}, \bibinfo {author} {\bibfnamefont {M.~W.}\ \bibnamefont
				{Johnson}}, \bibinfo {author} {\bibfnamefont {M.~H.}\ \bibnamefont {Amin}},
			\bibinfo {author} {\bibfnamefont {R.}~\bibnamefont {Harris}}, \bibinfo
			{author} {\bibfnamefont {F.}~\bibnamefont {Altomare}}, \bibinfo {author}
			{\bibfnamefont {A.~J.}\ \bibnamefont {Berkley}}, \bibinfo {author}
			{\bibfnamefont {P.}~\bibnamefont {Bunyk}}, \bibinfo {author} {\bibfnamefont
				{J.}~\bibnamefont {Cai}}, \bibinfo {author} {\bibfnamefont {E.~M.}\
				\bibnamefont {Chapple}}, \bibinfo {author} {\bibfnamefont {P.}~\bibnamefont
				{Chavez}}, \bibinfo {author} {\bibfnamefont {F.}~\bibnamefont {Cioata}},
			\bibinfo {author} {\bibfnamefont {T.}~\bibnamefont {Cirip}}, \bibinfo
			{author} {\bibfnamefont {P.}~\bibnamefont {deBuen}}, \bibinfo {author}
			{\bibfnamefont {M.}~\bibnamefont {Drew-Brook}}, \bibinfo {author}
			{\bibfnamefont {C.}~\bibnamefont {Enderud}}, \bibinfo {author} {\bibfnamefont
				{S.}~\bibnamefont {Gildert}}, \bibinfo {author} {\bibfnamefont
				{F.}~\bibnamefont {Hamze}}, \bibinfo {author} {\bibfnamefont {J.~P.}\
				\bibnamefont {Hilton}}, \bibinfo {author} {\bibfnamefont {E.}~\bibnamefont
				{Hoskinson}}, \bibinfo {author} {\bibfnamefont {K.}~\bibnamefont {Karimi}},
			\bibinfo {author} {\bibfnamefont {E.}~\bibnamefont {Ladizinsky}}, \bibinfo
			{author} {\bibfnamefont {N.}~\bibnamefont {Ladizinsky}}, \bibinfo {author}
			{\bibfnamefont {T.}~\bibnamefont {Lanting}}, \bibinfo {author} {\bibfnamefont
				{T.}~\bibnamefont {Mahon}}, \bibinfo {author} {\bibfnamefont
				{R.}~\bibnamefont {Neufeld}}, \bibinfo {author} {\bibfnamefont
				{T.}~\bibnamefont {Oh}}, \bibinfo {author} {\bibfnamefont {I.}~\bibnamefont
				{Perminov}}, \bibinfo {author} {\bibfnamefont {C.}~\bibnamefont {Petroff}},
			\bibinfo {author} {\bibfnamefont {A.}~\bibnamefont {Przybysz}}, \bibinfo
			{author} {\bibfnamefont {C.}~\bibnamefont {Rich}}, \bibinfo {author}
			{\bibfnamefont {P.}~\bibnamefont {Spear}}, \bibinfo {author} {\bibfnamefont
				{A.}~\bibnamefont {Tcaciuc}}, \bibinfo {author} {\bibfnamefont {M.~C.}\
				\bibnamefont {Thom}}, \bibinfo {author} {\bibfnamefont {E.}~\bibnamefont
				{Tolkacheva}}, \bibinfo {author} {\bibfnamefont {S.}~\bibnamefont
				{Uchaikin}}, \bibinfo {author} {\bibfnamefont {J.}~\bibnamefont {Wang}},
			\bibinfo {author} {\bibfnamefont {A.~B.}\ \bibnamefont {Wilson}}, \bibinfo
			{author} {\bibfnamefont {Z.}~\bibnamefont {Merali}}, \ and\ \bibinfo {author}
			{\bibfnamefont {G.}~\bibnamefont {Rose}},\ }\href {\doibase
			10.1038/ncomms2920} {\bibfield  {journal} {\bibinfo  {journal} {Nature
					Communications}\ }\textbf {\bibinfo {volume} {4}} (\bibinfo {year} {2013}),\
			10.1038/ncomms2920}\BibitemShut {NoStop}%
		\bibitem [{\citenamefont {Arceci}\ \emph {et~al.}(2017)\citenamefont {Arceci},
			\citenamefont {Barbarino}, \citenamefont {Fazio},\ and\ \citenamefont
			{Santoro}}]{arceci:dissipative-lz}%
		\BibitemOpen
		\bibfield  {author} {\bibinfo {author} {\bibfnamefont {L.}~\bibnamefont
				{Arceci}}, \bibinfo {author} {\bibfnamefont {S.}~\bibnamefont {Barbarino}},
			\bibinfo {author} {\bibfnamefont {R.}~\bibnamefont {Fazio}}, \ and\ \bibinfo
			{author} {\bibfnamefont {G.~E.}\ \bibnamefont {Santoro}},\ }\href {\doibase
			10.1103/PhysRevB.96.054301} {\bibfield  {journal} {\bibinfo  {journal} {Phys.
					Rev. B}\ }\textbf {\bibinfo {volume} {96}},\ \bibinfo {pages} {054301}
			(\bibinfo {year} {2017})}\BibitemShut {NoStop}%
		\bibitem [{\citenamefont {Cialdi}\ \emph {et~al.}(2011)\citenamefont {Cialdi},
			\citenamefont {Brivio}, \citenamefont {Tesio},\ and\ \citenamefont
			{Paris}}]{CialdiExp}%
		\BibitemOpen
		\bibfield  {author} {\bibinfo {author} {\bibfnamefont {S.}~\bibnamefont
				{Cialdi}}, \bibinfo {author} {\bibfnamefont {D.}~\bibnamefont {Brivio}},
			\bibinfo {author} {\bibfnamefont {E.}~\bibnamefont {Tesio}}, \ and\ \bibinfo
			{author} {\bibfnamefont {M.~G.~A.}\ \bibnamefont {Paris}},\ }\href {\doibase
			10.1103/PhysRevA.83.042308} {\bibfield  {journal} {\bibinfo  {journal} {Phys.
					Rev. A}\ }\textbf {\bibinfo {volume} {83}},\ \bibinfo {pages} {042308}
			(\bibinfo {year} {2011})}\BibitemShut {NoStop}%
		\bibitem [{\citenamefont {Liu}\ \emph {et~al.}(2011)\citenamefont {Liu},
			\citenamefont {Li}, \citenamefont {Huang}, \citenamefont {Li}, \citenamefont
			{Guo}, \citenamefont {Laine}, \citenamefont {Breuer},\ and\ \citenamefont
			{Piilo}}]{NatureChina}%
		\BibitemOpen
		\bibfield  {author} {\bibinfo {author} {\bibfnamefont {B.-H.}\ \bibnamefont
				{Liu}}, \bibinfo {author} {\bibfnamefont {L.}~\bibnamefont {Li}}, \bibinfo
			{author} {\bibfnamefont {Y.-F.}\ \bibnamefont {Huang}}, \bibinfo {author}
			{\bibfnamefont {C.-F.}\ \bibnamefont {Li}}, \bibinfo {author} {\bibfnamefont
				{G.-C.}\ \bibnamefont {Guo}}, \bibinfo {author} {\bibfnamefont {E.~M.}\
				\bibnamefont {Laine}}, \bibinfo {author} {\bibfnamefont {H.~P.}\ \bibnamefont
				{Breuer}}, \ and\ \bibinfo {author} {\bibfnamefont {J.}~\bibnamefont
				{Piilo}},\ }\href {\doibase 10.1038/nphys2085} {\bibfield  {journal}
			{\bibinfo  {journal} {Nature Physics}\ }\textbf {\bibinfo {volume} {7}},\
			\bibinfo {pages} {931} (\bibinfo {year} {2011})}\BibitemShut {NoStop}%
		\bibitem [{\citenamefont {Thorwart}\ \emph {et~al.}(2005)\citenamefont
			{Thorwart}, \citenamefont {Eckel},\ and\ \citenamefont
			{Mucciolo}}]{thorwart3}%
		\BibitemOpen
		\bibfield  {author} {\bibinfo {author} {\bibfnamefont {M.}~\bibnamefont
				{Thorwart}}, \bibinfo {author} {\bibfnamefont {J.}~\bibnamefont {Eckel}}, \
			and\ \bibinfo {author} {\bibfnamefont {E.~R.}\ \bibnamefont {Mucciolo}},\
		}\href {\doibase 10.1103/PhysRevB.72.235320} {\bibfield  {journal} {\bibinfo
				{journal} {Phys. Rev. B}\ }\textbf {\bibinfo {volume} {72}},\ \bibinfo
			{pages} {235320} (\bibinfo {year} {2005})}\BibitemShut {NoStop}%
		\bibitem [{\citenamefont {Goychuk}\ and\ \citenamefont
			{H\"anggi}(2006)}]{goychuk:2006}%
		\BibitemOpen
		\bibfield  {author} {\bibinfo {author} {\bibfnamefont {I.}~\bibnamefont
				{Goychuk}}\ and\ \bibinfo {author} {\bibfnamefont {P.}~\bibnamefont
				{H\"anggi}},\ }\href {\doibase
			https://doi.org/10.1016/j.chemphys.2005.11.026} {\bibfield  {journal}
			{\bibinfo  {journal} {Chemical Physics}\ }\textbf {\bibinfo {volume} {324}},\
			\bibinfo {pages} {160} (\bibinfo {year} {2006})}\BibitemShut {NoStop}%
		\bibitem [{\citenamefont {de~Vega}\ \emph {et~al.}(2015)\citenamefont
			{de~Vega}, \citenamefont {Schollw\"ock},\ and\ \citenamefont
			{Wolf}}]{de-vega:discretize-baths}%
		\BibitemOpen
		\bibfield  {author} {\bibinfo {author} {\bibfnamefont {I.}~\bibnamefont
				{de~Vega}}, \bibinfo {author} {\bibfnamefont {U.}~\bibnamefont
				{Schollw\"ock}}, \ and\ \bibinfo {author} {\bibfnamefont {F.~A.}\
				\bibnamefont {Wolf}},\ }\href {\doibase 10.1103/PhysRevB.92.155126}
		{\bibfield  {journal} {\bibinfo  {journal} {Phys. Rev. B}\ }\textbf {\bibinfo
				{volume} {92}},\ \bibinfo {pages} {155126} (\bibinfo {year}
			{2015})}\BibitemShut {NoStop}%
		\bibitem [{\citenamefont {Park}\ and\ \citenamefont {Light}(1986)}]{Lanczos1}%
		\BibitemOpen
		\bibfield  {author} {\bibinfo {author} {\bibfnamefont {T.~J.}\ \bibnamefont
				{Park}}\ and\ \bibinfo {author} {\bibfnamefont {J.~C.}\ \bibnamefont
				{Light}},\ }\href {\doibase 10.1063/1.451548} {\bibfield  {journal} {\bibinfo
				{journal} {The Journal of Chemical Physics}\ }\textbf {\bibinfo {volume}
				{85}},\ \bibinfo {pages} {5870} (\bibinfo {year} {1986})},\ \Eprint
		{http://arxiv.org/abs/https://doi.org/10.1063/1.451548}
		{https://doi.org/10.1063/1.451548} \BibitemShut {NoStop}%
		\bibitem [{\citenamefont {Frapiccini}\ \emph {et~al.}(2014)\citenamefont
			{Frapiccini}, \citenamefont {Hamido}, \citenamefont {Schr\"oter},
			\citenamefont {Pyke}, \citenamefont {Mota-Furtado}, \citenamefont {O'Mahony},
			\citenamefont {Madro\~nero}, \citenamefont {Eiglsperger},\ and\ \citenamefont
			{Piraux}}]{Lanczos2}%
		\BibitemOpen
		\bibfield  {author} {\bibinfo {author} {\bibfnamefont {A.~L.}\ \bibnamefont
				{Frapiccini}}, \bibinfo {author} {\bibfnamefont {A.}~\bibnamefont {Hamido}},
			\bibinfo {author} {\bibfnamefont {S.}~\bibnamefont {Schr\"oter}}, \bibinfo
			{author} {\bibfnamefont {D.}~\bibnamefont {Pyke}}, \bibinfo {author}
			{\bibfnamefont {F.}~\bibnamefont {Mota-Furtado}}, \bibinfo {author}
			{\bibfnamefont {P.~F.}\ \bibnamefont {O'Mahony}}, \bibinfo {author}
			{\bibfnamefont {J.}~\bibnamefont {Madro\~nero}}, \bibinfo {author}
			{\bibfnamefont {J.}~\bibnamefont {Eiglsperger}}, \ and\ \bibinfo {author}
			{\bibfnamefont {B.}~\bibnamefont {Piraux}},\ }\href {\doibase
			10.1103/PhysRevA.89.023418} {\bibfield  {journal} {\bibinfo  {journal} {Phys.
					Rev. A}\ }\textbf {\bibinfo {volume} {89}},\ \bibinfo {pages} {023418}
			(\bibinfo {year} {2014})}\BibitemShut {NoStop}%
		\bibitem [{\citenamefont {Novelli}\ \emph {et~al.}(2014)\citenamefont
			{Novelli}, \citenamefont {De~Filippis}, \citenamefont {Cataudella},
			\citenamefont {Esposito}, \citenamefont {Vergara}, \citenamefont {Cilento},
			\citenamefont {Sindici}, \citenamefont {Amaricci}, \citenamefont {Giannetti},
			\citenamefont {Prabhakaran}, \citenamefont {Wall}, \citenamefont {Perucchi},
			\citenamefont {Dal~Conte}, \citenamefont {Cerullo}, \citenamefont {Capone},
			\citenamefont {Mishchenko}, \citenamefont {Gr\"{u}ninger}, \citenamefont
			{Nagaosa}, \citenamefont {Parmigiani},\ and\ \citenamefont
			{Fausti}}]{nature:giulio}%
		\BibitemOpen
		\bibfield  {author} {\bibinfo {author} {\bibfnamefont {F.}~\bibnamefont
				{Novelli}}, \bibinfo {author} {\bibfnamefont {G.}~\bibnamefont
				{De~Filippis}}, \bibinfo {author} {\bibfnamefont {V.}~\bibnamefont
				{Cataudella}}, \bibinfo {author} {\bibfnamefont {M.}~\bibnamefont
				{Esposito}}, \bibinfo {author} {\bibfnamefont {I.}~\bibnamefont {Vergara}},
			\bibinfo {author} {\bibfnamefont {F.}~\bibnamefont {Cilento}}, \bibinfo
			{author} {\bibfnamefont {E.}~\bibnamefont {Sindici}}, \bibinfo {author}
			{\bibfnamefont {A.}~\bibnamefont {Amaricci}}, \bibinfo {author}
			{\bibfnamefont {C.}~\bibnamefont {Giannetti}}, \bibinfo {author}
			{\bibfnamefont {D.}~\bibnamefont {Prabhakaran}}, \bibinfo {author}
			{\bibfnamefont {S.}~\bibnamefont {Wall}}, \bibinfo {author} {\bibfnamefont
				{A.}~\bibnamefont {Perucchi}}, \bibinfo {author} {\bibfnamefont
				{S.}~\bibnamefont {Dal~Conte}}, \bibinfo {author} {\bibfnamefont
				{G.}~\bibnamefont {Cerullo}}, \bibinfo {author} {\bibfnamefont
				{M.}~\bibnamefont {Capone}}, \bibinfo {author} {\bibfnamefont
				{A.}~\bibnamefont {Mishchenko}}, \bibinfo {author} {\bibfnamefont
				{M.}~\bibnamefont {Gr\"{u}ninger}}, \bibinfo {author} {\bibfnamefont
				{N.}~\bibnamefont {Nagaosa}}, \bibinfo {author} {\bibfnamefont
				{F.}~\bibnamefont {Parmigiani}}, \ and\ \bibinfo {author} {\bibfnamefont
				{D.}~\bibnamefont {Fausti}},\ }\href {\doibase 10.1038/ncomms6112} {\bibfield
			{journal} {\bibinfo  {journal} {Nature Communications}\ }\textbf {\bibinfo
				{volume} {5}} (\bibinfo {year} {2014}),\ 10.1038/ncomms6112}\BibitemShut
		{NoStop}%
		\bibitem [{\citenamefont {De~Filippis}\ \emph
			{et~al.}(2012{\natexlab{a}})\citenamefont {De~Filippis}, \citenamefont
			{Cataudella}, \citenamefont {Nowadnick}, \citenamefont {Devereaux},
			\citenamefont {Mishchenko},\ and\ \citenamefont {Nagaosa}}]{letters:giulio1}%
		\BibitemOpen
		\bibfield  {author} {\bibinfo {author} {\bibfnamefont {G.}~\bibnamefont
				{De~Filippis}}, \bibinfo {author} {\bibfnamefont {V.}~\bibnamefont
				{Cataudella}}, \bibinfo {author} {\bibfnamefont {E.~A.}\ \bibnamefont
				{Nowadnick}}, \bibinfo {author} {\bibfnamefont {T.~P.}\ \bibnamefont
				{Devereaux}}, \bibinfo {author} {\bibfnamefont {A.~S.}\ \bibnamefont
				{Mishchenko}}, \ and\ \bibinfo {author} {\bibfnamefont {N.}~\bibnamefont
				{Nagaosa}},\ }\href {\doibase 10.1103/PhysRevLett.109.176402} {\bibfield
			{journal} {\bibinfo  {journal} {Phys. Rev. Lett.}\ }\textbf {\bibinfo
				{volume} {109}},\ \bibinfo {pages} {176402} (\bibinfo {year}
			{2012}{\natexlab{a}})}\BibitemShut {NoStop}%
		\bibitem [{\citenamefont {De~Filippis}\ \emph
			{et~al.}(2012{\natexlab{b}})\citenamefont {De~Filippis}, \citenamefont
			{Cataudella}, \citenamefont {Mishchenko},\ and\ \citenamefont
			{Nagaosa}}]{prb:giulio}%
		\BibitemOpen
		\bibfield  {author} {\bibinfo {author} {\bibfnamefont {G.}~\bibnamefont
				{De~Filippis}}, \bibinfo {author} {\bibfnamefont {V.}~\bibnamefont
				{Cataudella}}, \bibinfo {author} {\bibfnamefont {A.~S.}\ \bibnamefont
				{Mishchenko}}, \ and\ \bibinfo {author} {\bibfnamefont {N.}~\bibnamefont
				{Nagaosa}},\ }\href {\doibase 10.1103/PhysRevB.85.094302} {\bibfield
			{journal} {\bibinfo  {journal} {Phys. Rev. B}\ }\textbf {\bibinfo {volume}
				{85}},\ \bibinfo {pages} {094302} (\bibinfo {year}
			{2012}{\natexlab{b}})}\BibitemShut {NoStop}%
		\bibitem [{\citenamefont {Marchand}\ \emph {et~al.}(2010)\citenamefont
			{Marchand}, \citenamefont {De~Filippis}, \citenamefont {Cataudella},
			\citenamefont {Berciu}, \citenamefont {Nagaosa}, \citenamefont {Prokof'ev},
			\citenamefont {Mishchenko},\ and\ \citenamefont {Stamp}}]{letters:giulio2}%
		\BibitemOpen
		\bibfield  {author} {\bibinfo {author} {\bibfnamefont {D.~J.~J.}\
				\bibnamefont {Marchand}}, \bibinfo {author} {\bibfnamefont {G.}~\bibnamefont
				{De~Filippis}}, \bibinfo {author} {\bibfnamefont {V.}~\bibnamefont
				{Cataudella}}, \bibinfo {author} {\bibfnamefont {M.}~\bibnamefont {Berciu}},
			\bibinfo {author} {\bibfnamefont {N.}~\bibnamefont {Nagaosa}}, \bibinfo
			{author} {\bibfnamefont {N.~V.}\ \bibnamefont {Prokof'ev}}, \bibinfo {author}
			{\bibfnamefont {A.~S.}\ \bibnamefont {Mishchenko}}, \ and\ \bibinfo {author}
			{\bibfnamefont {P.~C.~E.}\ \bibnamefont {Stamp}},\ }\href {\doibase
			10.1103/PhysRevLett.105.266605} {\bibfield  {journal} {\bibinfo  {journal}
				{Phys. Rev. Lett.}\ }\textbf {\bibinfo {volume} {105}},\ \bibinfo {pages}
			{266605} (\bibinfo {year} {2010})}\BibitemShut {NoStop}%
		\bibitem [{\citenamefont {Bulla}\ \emph {et~al.}(2003)\citenamefont {Bulla},
			\citenamefont {Tong},\ and\ \citenamefont {Vojta}}]{bulla:nrg1}%
		\BibitemOpen
		\bibfield  {author} {\bibinfo {author} {\bibfnamefont {R.}~\bibnamefont
				{Bulla}}, \bibinfo {author} {\bibfnamefont {N.-H.}\ \bibnamefont {Tong}}, \
			and\ \bibinfo {author} {\bibfnamefont {M.}~\bibnamefont {Vojta}},\ }\href
		{\doibase 10.1103/PhysRevLett.91.170601} {\bibfield  {journal} {\bibinfo
				{journal} {Phys. Rev. Lett.}\ }\textbf {\bibinfo {volume} {91}},\ \bibinfo
			{pages} {170601} (\bibinfo {year} {2003})}\BibitemShut {NoStop}%
		\bibitem [{\citenamefont {Vojta}\ \emph {et~al.}(2005)\citenamefont {Vojta},
			\citenamefont {Tong},\ and\ \citenamefont {Bulla}}]{bulla:nrg2}%
		\BibitemOpen
		\bibfield  {author} {\bibinfo {author} {\bibfnamefont {M.}~\bibnamefont
				{Vojta}}, \bibinfo {author} {\bibfnamefont {N.-H.}\ \bibnamefont {Tong}}, \
			and\ \bibinfo {author} {\bibfnamefont {R.}~\bibnamefont {Bulla}},\ }\href
		{\doibase 10.1103/PhysRevLett.94.070604} {\bibfield  {journal} {\bibinfo
				{journal} {Phys. Rev. Lett.}\ }\textbf {\bibinfo {volume} {94}},\ \bibinfo
			{pages} {070604} (\bibinfo {year} {2005})}\BibitemShut {NoStop}%
		\bibitem [{\citenamefont {Zhang}\ \emph {et~al.}(2010)\citenamefont {Zhang},
			\citenamefont {Chen},\ and\ \citenamefont {Wang}}]{zhang:sub-ohm-sbm}%
		\BibitemOpen
		\bibfield  {author} {\bibinfo {author} {\bibfnamefont {Y.-Y.}\ \bibnamefont
				{Zhang}}, \bibinfo {author} {\bibfnamefont {Q.-H.}\ \bibnamefont {Chen}}, \
			and\ \bibinfo {author} {\bibfnamefont {K.-L.}\ \bibnamefont {Wang}},\ }\href
		{\doibase 10.1103/PhysRevB.81.121105} {\bibfield  {journal} {\bibinfo
				{journal} {Phys. Rev. B}\ }\textbf {\bibinfo {volume} {81}},\ \bibinfo
			{pages} {121105} (\bibinfo {year} {2010})}\BibitemShut {NoStop}%
		\bibitem [{\citenamefont {Mahan}(2000)}]{mahan:many-particle}%
		\BibitemOpen
		\bibfield  {author} {\bibinfo {author} {\bibfnamefont {G.~D.}\ \bibnamefont
				{Mahan}},\ }\href@noop {} {\emph {\bibinfo {title} {Many-particle
					physics}}},\ \bibinfo {edition} {3rd}\ ed.\ (\bibinfo  {publisher}
		{Springer},\ \bibinfo {year} {2000})\BibitemShut {NoStop}%
		\bibitem [{\citenamefont {Albash}\ and\ \citenamefont
			{Lidar}(2015)}]{albash:decoherence}%
		\BibitemOpen
		\bibfield  {author} {\bibinfo {author} {\bibfnamefont {T.}~\bibnamefont
				{Albash}}\ and\ \bibinfo {author} {\bibfnamefont {D.~A.}\ \bibnamefont
				{Lidar}},\ }\href {http://link.aps.org/doi/10.1103/PhysRevA.91.062320}
		{\bibfield  {journal} {\bibinfo  {journal} {Phys. Rev. A}\ }\textbf {\bibinfo
				{volume} {91}},\ \bibinfo {pages} {062320} (\bibinfo {year}
			{2015})}\BibitemShut {NoStop}%
		\bibitem [{\citenamefont {Albash}\ \emph {et~al.}(2012)\citenamefont {Albash},
			\citenamefont {Boixo}, \citenamefont {Lidar},\ and\ \citenamefont
			{Zanardi}}]{Zanardi}%
		\BibitemOpen
		\bibfield  {author} {\bibinfo {author} {\bibfnamefont {T.}~\bibnamefont
				{Albash}}, \bibinfo {author} {\bibfnamefont {S.}~\bibnamefont {Boixo}},
			\bibinfo {author} {\bibfnamefont {D.~A.}\ \bibnamefont {Lidar}}, \ and\
			\bibinfo {author} {\bibfnamefont {P.}~\bibnamefont {Zanardi}},\ }\href
		{http://stacks.iop.org/1367-2630/14/i=12/a=123016} {\bibfield  {journal}
			{\bibinfo  {journal} {New Journal of Physics}\ }\textbf {\bibinfo {volume}
				{14}},\ \bibinfo {pages} {123016} (\bibinfo {year} {2012})}\BibitemShut
		{NoStop}%
		\bibitem [{\citenamefont {Feynman}\ and\ \citenamefont
			{Vernon}(1963)}]{FEYNMAN63}%
		\BibitemOpen
		\bibfield  {author} {\bibinfo {author} {\bibfnamefont {R.}~\bibnamefont
				{Feynman}}\ and\ \bibinfo {author} {\bibfnamefont {F.}~\bibnamefont
				{Vernon}},\ }\href {\doibase https://doi.org/10.1016/0003-4916(63)90068-X}
		{\bibfield  {journal} {\bibinfo  {journal} {Annals of Physics}\ }\textbf
			{\bibinfo {volume} {24}},\ \bibinfo {pages} {118 } (\bibinfo {year}
			{1963})}\BibitemShut {NoStop}%
		\bibitem [{\citenamefont {Orth}\ \emph {et~al.}(2013)\citenamefont {Orth},
			\citenamefont {Imambekov},\ and\ \citenamefont {Le~Hur}}]{orth-lehur}%
		\BibitemOpen
		\bibfield  {author} {\bibinfo {author} {\bibfnamefont {P.~P.}\ \bibnamefont
				{Orth}}, \bibinfo {author} {\bibfnamefont {A.}~\bibnamefont {Imambekov}}, \
			and\ \bibinfo {author} {\bibfnamefont {K.}~\bibnamefont {Le~Hur}},\ }\href
		{\doibase 10.1103/PhysRevB.87.014305} {\bibfield  {journal} {\bibinfo
				{journal} {Phys. Rev. B}\ }\textbf {\bibinfo {volume} {87}},\ \bibinfo
			{pages} {014305} (\bibinfo {year} {2013})}\BibitemShut {NoStop}%
		\bibitem [{\citenamefont {De~Filippis}\ and\ \citenamefont
			{De~Candia}()}]{DeFilippisunpub}%
		\BibitemOpen
		\bibfield  {author} {\bibinfo {author} {\bibfnamefont {G.}~\bibnamefont
				{De~Filippis}}\ and\ \bibinfo {author} {\bibfnamefont {A.}~\bibnamefont
				{De~Candia}},\ }\href@noop {} {\ }\bibinfo {note} {Unpublished}\BibitemShut
		{NoStop}%
		\bibitem [{\citenamefont {Landau}(1932)}]{landau:crossings}%
		\BibitemOpen
		\bibfield  {author} {\bibinfo {author} {\bibfnamefont {L.}~\bibnamefont
				{Landau}},\ }\href@noop {} {\bibfield  {journal} {\bibinfo  {journal}
				{Physikalische Zeitschrift der Sowjetunion}\ }\textbf {\bibinfo {volume}
				{2}},\ \bibinfo {pages} {46} (\bibinfo {year} {1932})}\BibitemShut {NoStop}%
		\bibitem [{\citenamefont {Zener}(1932)}]{zener:crossings}%
		\BibitemOpen
		\bibfield  {author} {\bibinfo {author} {\bibfnamefont {C.}~\bibnamefont
				{Zener}},\ }\href
		{http://rspa.royalsocietypublishing.org/content/137/833/696} {\bibfield
			{journal} {\bibinfo  {journal} {Proc. R. Soc. London, Ser. A}\ }\textbf
			{\bibinfo {volume} {137}},\ \bibinfo {pages} {696} (\bibinfo {year}
			{1932})}\BibitemShut {NoStop}%
		\bibitem [{\citenamefont {Nalbach}\ and\ \citenamefont
			{Thorwart}(2009)}]{thorwart1}%
		\BibitemOpen
		\bibfield  {author} {\bibinfo {author} {\bibfnamefont {P.}~\bibnamefont
				{Nalbach}}\ and\ \bibinfo {author} {\bibfnamefont {M.}~\bibnamefont
				{Thorwart}},\ }\href {\doibase 10.1103/PhysRevLett.103.220401} {\bibfield
			{journal} {\bibinfo  {journal} {Phys. Rev. Lett.}\ }\textbf {\bibinfo
				{volume} {103}},\ \bibinfo {pages} {220401} (\bibinfo {year}
			{2009})}\BibitemShut {NoStop}%
		\bibitem [{\citenamefont {Javanbakht}\ \emph {et~al.}(2015)\citenamefont
			{Javanbakht}, \citenamefont {Nalbach},\ and\ \citenamefont
			{Thorwart}}]{thorwart4}%
		\BibitemOpen
		\bibfield  {author} {\bibinfo {author} {\bibfnamefont {S.}~\bibnamefont
				{Javanbakht}}, \bibinfo {author} {\bibfnamefont {P.}~\bibnamefont {Nalbach}},
			\ and\ \bibinfo {author} {\bibfnamefont {M.}~\bibnamefont {Thorwart}},\
		}\href {\doibase 10.1103/PhysRevA.91.052103} {\bibfield  {journal} {\bibinfo
				{journal} {Phys. Rev. A}\ }\textbf {\bibinfo {volume} {91}},\ \bibinfo
			{pages} {052103} (\bibinfo {year} {2015})}\BibitemShut {NoStop}%
		\bibitem [{\citenamefont {Wubs}\ \emph {et~al.}(2006)\citenamefont {Wubs},
			\citenamefont {Saito}, \citenamefont {Kohler}, \citenamefont {H\"anggi},\
			and\ \citenamefont {Kayanuma}}]{wubs:lz-zero-t}%
		\BibitemOpen
		\bibfield  {author} {\bibinfo {author} {\bibfnamefont {M.}~\bibnamefont
				{Wubs}}, \bibinfo {author} {\bibfnamefont {K.}~\bibnamefont {Saito}},
			\bibinfo {author} {\bibfnamefont {S.}~\bibnamefont {Kohler}}, \bibinfo
			{author} {\bibfnamefont {P.}~\bibnamefont {H\"anggi}}, \ and\ \bibinfo
			{author} {\bibfnamefont {Y.}~\bibnamefont {Kayanuma}},\ }\href {\doibase
			10.1103/PhysRevLett.97.200404} {\bibfield  {journal} {\bibinfo  {journal}
				{Phys. Rev. Lett.}\ }\textbf {\bibinfo {volume} {97}},\ \bibinfo {pages}
			{200404} (\bibinfo {year} {2006})}\BibitemShut {NoStop}%
		\bibitem [{\citenamefont {Thorwart}\ \emph {et~al.}(2004)\citenamefont
			{Thorwart}, \citenamefont {Paladino},\ and\ \citenamefont
			{Grifoni}}]{thorwart2}%
		\BibitemOpen
		\bibfield  {author} {\bibinfo {author} {\bibfnamefont {M.}~\bibnamefont
				{Thorwart}}, \bibinfo {author} {\bibfnamefont {E.}~\bibnamefont {Paladino}},
			\ and\ \bibinfo {author} {\bibfnamefont {M.}~\bibnamefont {Grifoni}},\ }\href
		{\doibase https://doi.org/10.1016/j.chemphys.2003.10.007} {\bibfield
			{journal} {\bibinfo  {journal} {Chemical Physics}\ }\textbf {\bibinfo
				{volume} {296}},\ \bibinfo {pages} {333} (\bibinfo {year}
			{2004})}\BibitemShut {NoStop}%
		\bibitem [{\citenamefont {Huang}\ and\ \citenamefont
			{Zheng}(2008)}]{structured-bath-1}%
		\BibitemOpen
		\bibfield  {author} {\bibinfo {author} {\bibfnamefont {P.}~\bibnamefont
				{Huang}}\ and\ \bibinfo {author} {\bibfnamefont {H.}~\bibnamefont {Zheng}},\
		}\href {http://stacks.iop.org/0953-8984/20/i=39/a=395233} {\bibfield
			{journal} {\bibinfo  {journal} {Journal of Physics: Condensed Matter}\
			}\textbf {\bibinfo {volume} {20}},\ \bibinfo {pages} {395233} (\bibinfo
			{year} {2008})}\BibitemShut {NoStop}%
		\bibitem [{\citenamefont {Carrega}\ \emph {et~al.}(2016)\citenamefont
			{Carrega}, \citenamefont {Solinas}, \citenamefont {Sassetti},\ and\
			\citenamefont {Weiss}}]{structured-bath-2}%
		\BibitemOpen
		\bibfield  {author} {\bibinfo {author} {\bibfnamefont {M.}~\bibnamefont
				{Carrega}}, \bibinfo {author} {\bibfnamefont {P.}~\bibnamefont {Solinas}},
			\bibinfo {author} {\bibfnamefont {M.}~\bibnamefont {Sassetti}}, \ and\
			\bibinfo {author} {\bibfnamefont {U.}~\bibnamefont {Weiss}},\ }\href
		{\doibase 10.1103/PhysRevLett.116.240403} {\bibfield  {journal} {\bibinfo
				{journal} {Phys. Rev. Lett.}\ }\textbf {\bibinfo {volume} {116}},\ \bibinfo
			{pages} {240403} (\bibinfo {year} {2016})}\BibitemShut {NoStop}%
	\end{thebibliography}
	
	%

\end{document}